\documentclass[english,jfm]{jfm}
\usepackage[T1]{fontenc}
\usepackage[latin9]{inputenc}
\usepackage{verbatim}
\usepackage{amstext}
\usepackage{graphicx}
\usepackage{xfrac}
\usepackage{amssymb}
\usepackage[authoryear]{natbib}

\usepackage[colorinlistoftodos,prependcaption]{todonotes}

\graphicspath{{figures/}}

\usepackage{babel}

\begin{document}

\title[A statistical mechanics approach to mixing in stratified fluids]{A statistical mechanics approach to mixing in stratified fluids}

\author[A. Venaille, L. Gostiaux and J. Sommeria]%
{A.\ns V\ls E\ls N\ls A\ls I\ls L\ls L\ls E$^1$ \ns
\thanks{antoine.venaille@ens-lyon.fr},\ns
L.\ns  G\ls O\ls S\ls T\ls I\ls A\ls U\ls X$^2$ \and J.\ns S\ls O\ls M\ls M\ls E\ls R\ls I\ls A$^3$}

\affiliation{$^1$ Laboratoire de Physique UMR 5276 CNRS, ENS de Lyon, Universit\'e de Lyon, France, $^2$ LMFA UMR 5509 CNRS, Universit\'e de Lyon, France, $^3$ LEGI, CNRS, Universit\'e de Grenoble, France}

\date{\today}

\maketitle

\begin{abstract}
{  Predicting how much mixing occurs when a given amount of energy is injected into a Boussinesq fluid is a longstanding problem in stratified turbulence.
The huge number of degrees of freedom involved in these processes renders extremely difficult a deterministic approach to the problem.
Here we present a statistical mechanics approach yielding a prediction for a cumulative, global mixing efficiency as a function of a global Richardson number and the background buoyancy profile.
Assuming random evolution through turbulent stirring, the theory predicts that the inviscid, adiabatic dynamics is attracted irreversibly towards an equilibrium state characterised by a smooth, stable buoyancy profile at a coarse-grained level,  upon which are fine-scale fluctuations of velocity and buoyancy.
The convergence towards a coarse-grained buoyancy profile different from the initial one corresponds to an irreversible increase of potential energy, and the efficiency of mixing is quantified as the ratio of this potential energy increase to the total energy injected into the system. The remaining part of the energy is literally lost into small-scale fluctuations. 
}
We show that for sufficiently large Richardson number, there is equipartition between potential and kinetic energy, provided that the background buoyancy profile is strictly monotonic.
This yields a mixing efficiency of 0.25, which provides statistical mechanics support for previous predictions based on phenomenological kinematics arguments.  
In the general case, the cumulative, global  mixing efficiency predicted by the equilibrium theory can be computed using an algorithm based on a maximum entropy production principle.
It is shown in particular that the variation of mixing efficiency with the Richardson number strongly depends on the background buoyancy profile.
This approach {  could be} useful to the understanding of mixing in stratified turbulence in the limit of large Reynolds and P\'eclet numbers.

\end{abstract}

\section{Introduction}

\noindent

The large-scale stratification and dynamics of the oceans depend crucially on localised turbulent mixing events \citep{wunsch2004vertical,thorpe2005turbulent}.
These mixing processes occur on temporal and spatial scales much smaller than the current resolutions of general circulation models and must therefore be parameterised \citep{large1994oceanic}.
It is essential for that purpose to know how much mixing occurs when stratification is stirred by a turbulent flow \citep{hopfinger1987turbulence,fernando91,staquet2002internal,peltier2003,ivey2008}.
More precisely, which fraction of the injected energy is lost through a direct turbulent kinetic energy cascade and viscous dissipation, which fraction contributes to modifying the background stratification, and what is  the resulting vertical buoyancy profile ? Here we propose to use statistical mechanics as a guideline for the understanding of turbulent stirring and mixing in a stratified fluid.

Equilibrium statistical mechanics counts the available states of the system with given constraints based on conservation laws. Under random evolution, the system is expected to reach the \emph{macroscopic} state, which corresponds to the maximum number of \emph{microscopic} configurations. In this paper, the \emph{macroscopic} quantity to be determined by the theory is the partition between kinetic and potential energy, as well as the corresponding mean (coarse-grained) vertical buoyancy profile. The \emph{microscopic} configurations will consists of the set of all possible buoyancy fields and non-divergent velocity fields, and the constraints will be provided by dynamical invariants of the inviscid and adiabatic fluid.
 
The application of equilibrium statistical mechanics theory to systems described by continuous fields is however problematic; see e.g. \cite{Pomeau94}. Indeed, such systems are characterised by an infinite number of degrees of freedom, which can lead to an accumulation of energy at small scales, whose divergence can only be avoided by an artificial truncation in Fourier space.  \cite{kraichnan1967}  has however explained the energy cascade toward small scales as a trend of the system to approach  such equilibrium. By contrast, in two-dimensional turbulence, statistical equilibrium rather accumulates energy at large scale, which Kraichnan has related to the occurrence of an inverse energy cascade. The statistical equilibrium therefore reveals the trend of the evolution for the actual irreversible turbulent system {  in the limit of small viscosity}. We here follow a similar idea to study mixing in stratified fluids, using however a significantly different statistical mechanics approach.

Instead of considering Galerkin-truncated flows,  \cite{onsager1949} modelled the fluid continuum by a very large but finite set of singular point vortices to explain the self-organisation of  two-dimensional turbulent flows as a tendency to reach an equilibrium state, see also \cite{eyink2006onsager}.  Extensions of those ideas  to the  continuous two-dimensional Euler and quasi-geostrophic dynamics have been developed independently by \cite{miller1990} and \cite{robert1991} (MRS hereafter). {  A similar theory had been previously applied to the Vlasov equations by~\citep{lynden1967statistical} in order to predict self-organisation in plasma and self-gravitating systems, see e.g. \cite{chavanis2002statistical}}.  Subsequent work on the theoretical foundations of the approach, as well as on the analytical and numerical computation of equilibrium states is reviewed in \cite{sommeria2001two,majda2006,bouchet2012}. The theory introduces a truncation for the vorticity field, leading to unrealistic vorticity fluctuations at small scale, but it  provides quantitative predictions for the mean velocity field at large scale. In the geophysical context, the theory has been used { to explain  some features} of the structure of the Great Red  Spot of Jupiter \citep{turkington2001statistical,bouchet2002emergence}, oceanic rings and jets  \citep{weichman2006equilibrium,VenailleBouchetJPO10}, bottom-trapped oceanic recirculations \citep{venaille2012bottom}, the stratospheric polar vortex \citep{prieto2001analytical}, the vertical structure of geostrophic turbulence in stratified quasi-geostrophic turbulence \citep{merryfield1998effects,schecter2003maximum,VenailleVallisGriffies} and the structure of the thermocline in the global oceanic circulation \citep{salmon2012statistical}. {  One should however keep in mind that statistical equilibrium theory strictly applies to freely evolving flows, while most geophysical situations  involve forcing and friction. The equilibrium theory can be relevant to describe the large-scale flow when forcing and dissipation are sufficiently weak~\citep{majda2006,bouchet2009random}, but the values of conserved quantities are then set by the global balance between forcing and dissipation rather than from initial conditions.}

The equilibrium theory has already been derived for several flow systems that permit the existence of a direct energy cascade, such as three-dimensional axisymmetric Euler flows \citep{naso2010statistical,thalabard2014,Thalabard_ferroturb_NJP15}. 
The theory yields in that case predictions for the energy partition between toroidal and poloidal modes  \citep{thalabard2014}. Similarly, equilibrium theory has been used to predict the energy partition between inertia-gravity waves and vortical modes in shallow water models \citep{warn1986statistical,Weichman01,renaud2014}. Here we apply a similar approach to a non-rotating, density-stratified Boussinesq fluid in order to predict the partition between kinetic and potential energy for a given amount of energy injected into the system. 
\cite{tabak2004} computed the most probable buoyancy  field of a two-layer fluid with a prescribed total energy, assuming that the kinetic energy is constant at each height. Our contributions are twofold. First, we generalise their result to arbitrary buoyancy profiles, and obtain the kinetic energy profile as the output of the statistical theory. Second, we use these results to obtain  predictions for mixing efficiency in decaying configurations.  

{ How to infer the efficiency of mixing in forced-dissipative or decaying experiments has been carefully addressed in previous studies; see e.g. \cite{winters1995available,peltier2003,wykes2015meaning,salehipour2015diapycnal} and references therein. The traditional approach involves direct analyses of the diffusive destruction of small-scale density variance as the experiment proceeds, which in turn requires a separation of the influence of  stirring from that of irreversible mixing through application of the Lorenz concept of available potential energy that can be converted into kinetic energy and a base-state potential energy which can not. It has been demonstrated that the diffusive destruction of small-scale density variance may be represented by the time derivative $\mathcal{M}$ of base-state potential energy plus a small correction due to the action of molecular diffusion on the initial density stratification, a correction that becomes negligible in the limit of high Reynolds number \citep{winters1995available}. The time dependent efficiency of turbulent mixing may be then computed from the direct numerical simulations as $\eta_{t} = \mathcal{M}/( \mathcal{M}+\epsilon)$ where $\epsilon$ is the rate of viscous kinetic energy dissipation in the fluid domain \citep{peltier2003,salehipour2015diapycnal}. This definition of mixing efficiency is global in space since the computation of the base-state potential energy requires a rearrangement of the fluid particle at the domain scale.  Using a number of additional assumptions, it may be related to a local mixing efficiency that is often used in oceanography to model an effective diffusivity for diapycnal mixing \cite{osborn1980estimates,hopfinger1987turbulence,tailleux2009understanding}.  In decaying experiments, it is also convenient to define a cumulative mixing efficiency $\eta_{tot}= \int_0^{+\infty} \mathrm{d} t \mathcal{M} / \int_0^{+\infty} \mathrm{d} t \left(\mathcal{M} +\epsilon \right)$, which measures how much of the total injected energy has been used to irreversibly raise the potential energy of the system in the experiment. In practice, this quantity can easily be inferred in laboratory experiments by measuring the buoyancy profile once all dissipative effects have died-out, assuming the initial background stratification and the initial injected energy are known. 

{  Although the traditional approach to mixing efficiency in stratified turbulence emphasises the role of molecular diffusivity, we argue in this paper that irreversible mixing in decaying configurations can also be addressed within the framework of an inviscid, adiabatic Boussinesq flow model. Indeed, we will see that even if the background buoyancy field remains constant in time, the system is irreversibly attracted towards a state characterised by small-scale buoyancy fluctuations and a concomitant irreversible increase of available potential energy, assuming ergodicity.  This irreversibility is due to the fact that an overwhelming number of microscopic  configurations are close the most probable state, according to the equilibrium theory. The available potential energy of the equilibrium state could in principle be transferred back into kinetic energy, but this would correspond to the highly improbable escape from the equilibrium state. More precisely, we will show that the probability to observe a state different than the equilibrium state is vanishingly small (it tends to zero as the number of fluid particles tends to infinity). In other words, the so-called available potential energy of the system is statistically not available when the equilibrium state is reached, and we argue that a statistical mixing efficiency can be defined without reference to the molecular diffusion of buoyancy levels. Our working hypothesis is that this statistical mixing efficiency is equivalent to the traditional definition of mixing efficiency in the limit of weak molecular diffusivity.}

 Applying the statistical mechanics programs to Boussinesq  dynamics is done in three steps. The first step is to find relevant phase-space variables. These variables must satisfy a Liouville theorem, and we show in this paper (Appendix A) that this is the case of the velocity and buoyancy fields. This ensures that the dynamics is non-divergent in phase-space, so that the probability densities expressed in these variables  remain constant during the time evolution of the system. The fundamental postulate of equal probability for each microscopic configuration is then consistent with the dynamical evolution. Second, we need to introduce a discretisation of the continuous  fields describing the system. This technical step is classical when computing equilibrium states of systems described by deterministic partial differential equations.  Once the discrete approximation of the fields is introduced, one can count the  microscopic configurations, and the computation of the equilibrium states is rigorous. The third step is to introduce a macroscopic description of the system, and to find the most probable macrostates among all those that satisfy a set of  constraints provided by dynamical invariants. Using the equilibrium theory to describe the long time behaviour of the system finally requires  the assumption of ergodicity, i.e. that the system evenly explores phase space.  Even if the ergodic assumption may not be fulfilled in actual turbulent flow, computing the equilibria is at least a useful and necessary first step before addressing the out of equilibrium behaviour of the system in more comprehensive studies. 

 Denoting $H$ the height of the flow domain, $\Delta b$ the typical variations of the background buoyancy profile, $(U,\ L_t)$  the typical velocity and length scale of turbulence,  and $(\nu, \ \kappa)$ the molecular viscosity and diffusivity,  the efficiency of mixing depends \textit{a priori} on four non-dimensional parameters in laboratory or numerical experiments on stratified turbulence: a global Richardson number based on the domain scale $Ri=H\Delta b/U^2$, the Reynolds number $Re=UL_t/\nu$, the P\'eclet  number $Pe=UL_t/\kappa$,  and the ratio $L_t/H$ which depends on the energy injection mechanism.

The equilibrium statistical mechanics theory applies to the freely evolving inviscid adiabatic dynamics.  Considering such an approach to describe actual stratified turbulence amounts to assuming that the Reynolds number $Re$ and the P\'eclet number $Pe$ are sufficiently large, and that the typical time scale to approach the equilibrium state is smaller than the typical time scale for the dissipation of energy and buoyancy fluctuations.  Independently from statistical mechanics arguments, neglecting molecular effects is a natural assumption in the large Reynolds number limit, which has been proven useful in previous studies on three-dimensional turbulence \citep{eyink2006onsager}, in which case the observed dissipation rate of energy $\epsilon$ becomes independent from viscosity; see e.g. \cite{vassilicos2015dissipation} and references therein. 

{  Similarly, the independence of the dissipation rate of scalar fluctuations on the molecular diffusivity is a standard hypothesis in turbulence theory, coming back to the generalisation of Kolmogorov arguments by \cite{oboukhov1949structure} and \cite{corrsin1951spectrum}. In the case of a passive scalar, this hypothesis has been supported by experiments \citep{sreenivasan1996passive,warhaft2000passive}  and theoretical results \citep{shraiman2000scalar,falkovich2001particles}.}

Within the framework of the equilibrium theory, we assume conservation of the total energy and of the global distribution of buoyancy, but we show  that part of the energy and that part of the buoyancy fluctuations are irreversibly transferred to small scales once the equilibrium state is reached.  Since the amount of kinetic energy and buoyancy fluctuations that are irreversibly transferred to small scales can be computed explicitly within the equilibrium statistical mechanics framework, we argue that the theory makes possible a prediction for the cumulative mixing efficiency, even in the absence of viscosity or molecular dissipation in the model. Our working hypothesis is that those small scale fluctuations will be smoothed out by molecular effects over time scale much larger than the relaxation time towards equilibrium.} 
  


The paper is organised as follows. The equilibrium statistical mechanics theory is introduced and discussed in the second section. The actual computation of the equilibrium states is discussed in a third section. Application of the theory to predict mixing efficiency in freely-evolving flow (decaying turbulence) is discussed in a fourth section.{  We conclude and summarise the main results in the fifth section. Technical results on the Liouville theorem, on the computation of the macrostate entropy and on the numerical algorithm used to compute the equilibria are presented in two appendices.}

\section{Equilibrium statistical mechanics of non-rotating, density-stratified Boussinesq fluids}\label{sec:statmech}

\subsection{Dynamical system and invariants}

\noindent

We consider an inviscid  Boussinesq fluid that evolves in a three-dimensional  domain $\mathcal{V}_{\mathbf{x}}$  of volume $V$, see e.g. \cite{vallis2006}.
Spatial coordinates are denoted $\mathbf{x}=(x,y,z)$, with $\mathbf{e}_z$ the vertical unit vector pointing in the upward direction.
At each time $t$  the system is described by the buoyancy field $b=g \left(\varrho_0-\varrho \right)/\varrho_{0}$, where $\varrho(x,y,z,t)$ is the fluid density, $g$ gravity and $\varrho_0$ a reference density, and by the velocity field $\mathbf{u}=(u,v,w)$, which is non-divergent:
\begin{equation}
\nabla \cdot \mathbf{u}=0 \ .\label{eq:incompressibility}
\end{equation} 
In the absence of diffusivity, the buoyancy field is purely advected by the velocity field
\begin{equation}
\partial_{t} b +\mathbf{u}\cdot \nabla b = 0 \ , \label{eq:advection_buoyancy}
\end{equation}
and the dynamics of the velocity field is coupled to the buoyancy field through the momentum equation
\begin{equation}
\partial_{t}\mathbf{u}+\mathbf{u}\cdot \nabla\mathbf{u}=-\frac{1}{\varrho_0}\nabla P+b \mathbf{e}_z\ .\label{eq:momentum}
\end{equation}
Equation (\ref{eq:advection_buoyancy}) describes the Lagrangian conservation of the buoyancy. It implies the conservation of the global distribution (i.e. histogram) of buoyancy levels 
\begin{equation}
G(\sigma)=\frac{1}{V}\int_{\mathcal{V}_{\mathbf{x}}} \mathrm{d} \mathbf{x}  \  \delta(b-\sigma)\label{eq:histo}
\end{equation}
expressed as $\rm{d} G/\rm{d}t=0$. 
{  The conservation of $G(\sigma)$ is equivalent to the conservation of all the Casimir functionals $F[b]=\int \mathrm{d}\mathbf{x} \ f(b)$, with $f$ any arbitrary function; see e.g. \cite{potters2013sampling}. This conservation law is also equivalent to the conservation of the background (or sorted) buoyancy profile $b_s(s)$ defined as the buoyancy profile with minimal potential energy using
\begin{equation}
G(b_s) \mathrm{d} b_s = \frac{1}{2H} \mathrm{d} z. \label{eq:sorted}
\end{equation} 
}

Similarly, using Eqs. (\ref{eq:incompressibility}), (\ref{eq:advection_buoyancy}) and (\ref{eq:momentum}) one can show that  the total energy of the flow 
\begin{equation}
{E}=\int_{\mathcal{V}_{\mathbf{x}}} \mathrm{d}\mathbf{x}\ \left(\frac{1}{2}\mathbf{u}^{2}-bz\right) +\int_{\mathcal{V}_{\mathbf{x}}} \mathrm{d}\mathbf{x}\ z b_s \label{eq:ener}
\end{equation}
is another dynamical invariant{ : $\rm{d} E/\rm{d}t=0$}. Note that the total energy is defined up to a constant, but we have chosen this constant such that the energy vanishes  when there is no motion and when the buoyancy field is sorted ($E=0$ when $\mathbf{u}=0$ and $b=b_s$).

The Boussinesq equations are  characterised by additional dynamical invariants related to the conservation of Ertel potential vorticity, see e.g. \cite{salmon1998}. {  These invariants are essential to explain the occurrence of inverse cascade and self-organisation of the velocity field occurring in the presence of sufficiently large rotation. However various theoretical and numerical studies indicate that stratified turbulence in the absence of rotation is not influenced significantly by these invariants \citep{bartello1995geostrophic,lindborg2005effect,lindborg2006energy,waite2004stratified,herbert2014restricted}. We will therefore not consider the constraints related to the conservation of Ertel potential vorticity in the remaining of this paper. In the context of equilibrium statistical mechanics, this amounts to assuming that the entropy maxima obtained with and without these constraints are the same.}

\subsection{Microscopic configurations, macroscopic description and variational problem}

{ 

\noindent

For an isolated system, the fundamental postulate of equilibrium statistical mechanics is the equiprobability of the microscopic configurations corresponding to the same values of the dynamical invariants. 

The first step is to define what are the relevant phase-space variables describing these \textit{microscopic configurations}. Those variables must satisfy a Liouville's theorem, which means that the flow in phase space is non-divergent. This ensures that  microscopic configurations remain equiprobable during the time evolution of the system. We show in Appendix \ref{app:Liouville} that  the quadruplet of fields $(b,\mathbf{u})$ satisfy such a Liouvillle theorem, and are therefore relevant phase-space variables.

The second step is to identify the relevant dynamical invariants, which are here the total energy and the global distribution of buoyancy levels, defined in Eq. (\ref{eq:ener}) and in Eq. (\ref{eq:histo}), respectively. The ensemble of microscopic configurations characterised by the same dynamical invariants is called the microcanonical ensemble. This is the relevant ensemble to consider for an isolated system such as the unforced, inviscid, adiabatic Boussinesq system. 

The third step is to identify relevant macrostates, which describe an ensemble of microscopic configurations. We introduce for that purpose the probability $\rho(\mathbf{x},\sigma,\mathbf{v})$ of finding the buoyancy level $\sigma$ and the velocity level $\mathbf{v}$ in the vicinity of point $\mathbf{x}$. It is normalised at each point:  
\begin{equation}
\forall \mathbf{x} \in \mathcal{V}_{\mathbf{x}} ,\ \mathcal{N}_{\mathbf{x}}[\rho] =\int_{\mathcal{V}_{\mathbf{v}}}  \mathrm{d} \mathbf{v}\ \int_{\mathcal{V}_{\sigma}}  \mathrm{d} \sigma  \ \rho(\mathbf{x},\sigma,\mathbf{v}) =1 \ , \label{eq:norm}
\end{equation} 
where the integral bounds are 
\begin{equation}
{\mathcal{V}_{\mathbf{v}}}=\left[ -\infty, \ +\infty \right]^3 ,\quad {\mathcal{V}_{\sigma}}=\left[ -\infty, \ +\infty \right] . \label{eq:def_int_bounf}
\end{equation} 
Each microscopic state $\left(b(\mathbf{x}),\mathbf{u}(\mathbf{x})\right)$ is described at a macroscopic level by the PDF $\rho(\mathbf{x},\sigma,\mathbf{v})$, and many microscopic configurations are in general associated with a given PDF $\rho(\mathbf{x},\sigma,\mathbf{v})$, which is called a Young measure in mathematics; see e.g. \cite{robert1991}. 
%
%

Let us define more precisely how to compute the macroscopic state $\rho(\mathbf{x},\sigma,\mathbf{v})$ from a given microscopic configuration $(b(\mathbf{x}),\mathbf{u}(\mathbf{x}))$, which will be useful to count the number of microscopic configurations associated with a given macrostate.
For that purpose, we follow a procedure which is standard in the framework of equilibrium statistical mechanics of fluid systems, using a discrete approximation of the continuous fields. 
We consider a uniform \emph{coarse-grained} grid containing $N$ macrocells, and a \emph{fine-grained} grid obtained by dividing each macrocell of the coarse-grained grid into a uniform grid containing $M$ fluid particles, see  Fig. \ref{fig:schema}. 
On the one hand, discretisation of the \emph{microscopic} field $b(\mathbf{x})$ and $\mathbf{u}(\mathbf{x})$ are defined on the \emph{fine-grained} grid, which contains $MN$ fluid particles. 
This procedure also requires a discretisation of the buoyancy and velocity levels carried by the fluid particles, which is further discussed in Appendix  \ref{app:A}.
On the other hand,  the discrete approximation of the PDF $\rho$ is defined on the \textit{coarse-grained} grid.

For a given microscopic configuration, one can compute within each macrocell of the \textit{coarse-grained} grid the frequency of occurrence of buoyancy and velocity levels (a normalised histogram). 
In the limit $M\rightarrow +\infty$, for a prescribed value of $N$, the discrete approximations of the microscopic configurations tend to the continuous ones, and the discrete approximation of the PDF $\rho$ is equivalent to the frequency of occurrence of buoyancy and velocity levels within each macrocell. 
In other words, the discrete approximation of the PDF $\rho$ can be interpreted as the volume proportion of fluid particles carrying the buoyancy level $\sigma$ and velocity level $\mathbf{v}$ inside each macrocell. 
The continuous PDF field $\rho$ is then recovered by considering the limit $N\rightarrow +\infty$, which corresponds to the limit of infinitesimal macrocells. 
Several useful macroscopic quantities can be deduced from $\rho$, such as the macroscopic buoyancy field 
\begin{equation}
\overline{b}(\mathbf{x}) =  \int_{\mathcal{V}_{\sigma}}  \mathrm{d} \sigma \int_{\mathcal{V}_{{\mathbf{v}}}} \mathrm{d} \mathbf{v} \  \rho \sigma  \ , \label{eq:b_local_rho}
\end{equation}
and the local eddy kinetic energy field 
\begin{equation}
\frac{1}{2}\overline{\mathbf{u}^2}(\mathbf{x}) =  \int_{\mathcal{V}_{\sigma}}  \mathrm{d} \sigma \int_{\mathcal{V}_{{\mathbf{v}}}} \mathrm{d} \mathbf{v} \  \frac{1}{2} \rho  \mathbf{v}^{2} \ .\label{eq:KE_local_rho}
\end{equation}
Within the framework of the discrete approximation depicted in Fig. \ref{fig:schema}, those macroscopic quantities correspond to averages over macrocells, i.e. to a spatial coarse-graining at the scale of a macrocell $\sim N^{-1/3}$. Importantly, the small-scale fluctuations described by the macroscopic states are confined  at spatial scales below this coarse-graining scale, which tends to zero in the  limit $N\rightarrow +\infty$. %
 
\begin{figure}
\begin{center}
\includegraphics[width=\textwidth]{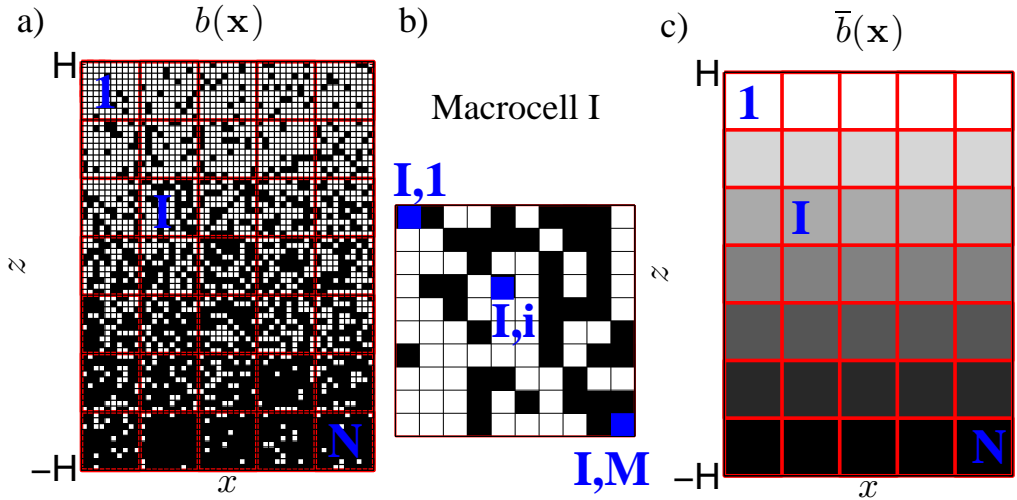}
\end{center}
\caption{a) A microscopic configuration of the discretised buoyancy field $b(\mathbf{x})$.  The discretised buoyancy field is defined on a uniform \textit{fine-grained} grid containing $M\times N$ elements, where $N$ is the number of grid points of the uniform \textit{coarse-grained} grid (red color).  b) Zoom on a single macrocell, containing $M$ microcells. Each microcell contains one fluid particle. Here we consider the case of a two-level system: the buoyancy carried by each fluid particle is $b=\pm \Delta b/2$. c) The macroscopic buoyancy field $\overline{b}({x})$ is defined on the uniform \textit{coarse-grained} grid (red colour), and is computed in the limit $M \rightarrow+\infty$ by averaging the microscopic buoyancy field within each macrocell, see e.g. \cite{miller1990,tabak2004}.\label{fig:schema}}
\end{figure}
 
The advantage of considering the probability field $\rho$  rather than only the coarse-grained fields such as $\overline{b}$ for a macroscopic description of the system is that global constraints provided by dynamical invariants can be expressed in terms of $\rho$.  The global constraints are given by the energy and the global distribution of buoyancy levels, which are defined as functional of phase-space variables $(\mathbf{u},b)$ in Eqs. (\ref{eq:ener}) and (\ref{eq:histo}),  respectively. Considering the discrete approximation described in the previous paragraph,  decomposing the spatial integrals appearing in  Eqs. (\ref{eq:ener}) and (\ref{eq:histo})  as a sum of spatial integrals over each macrocells, remembering then that the PDF $\rho$ is the frequency of occurrence of buoyancy and velocity levels within a given macrocell, and taking finally the limit $M\rightarrow+\infty$, $N\rightarrow +\infty$, the energy and the global distribution of buoyancy levels can be expressed as functionals of the PDF $\rho$:
\begin{equation}
\mathcal{E}[\rho]= \int_{\mathcal{V}_{\mathbf{x}}}   \mathrm{d}\mathbf{x} \int_{\mathcal{V}_{\mathbf{v}}}  \mathrm{d}\mathbf{v} \int_{\mathcal{V}_{\sigma}}  \mathrm{d}\sigma \ \rho\left(\frac{\mathbf{v}^{2}}{2}-\sigma z\right)   +\int_{\mathcal{V}_{\mathbf{x}}} \mathrm{d}\mathbf{x}\ z b_s\ ,\label{eq:E_rhop}
\end{equation}
\begin{equation}
\mathcal{G}_{\sigma}[\rho]=\int_{\mathcal{V}_{\mathbf{x}}}   \mathrm{d}\mathbf{x} \int_{\mathcal{V}_{\mathbf{v}}}  \mathrm{d}\mathbf{v} \  \rho \ .\label{eq:HHH}
\end{equation}
The microcanonical ensemble is defined by the ensemble of microstates characterised by the same energy $E$ and global distribution of buoyancy levels $G(\sigma)$. This ensemble contains therefore all the macroscopic states that satisfy the dynamical constraints $\mathcal{E}[\rho]=E$ and $\mathcal{G}_{\sigma}[\rho]=G(\sigma)$.

The last step is to count how many microscopic configurations are associated with a given macrostate.  Considering our discrete approximation of the fields, it is shown in Appendix \ref{app:A} that within the microcanonical ensemble, an overwhelming number of the microscopic configurations is concentrated close to the  most probable macrostate, which maximises the macrostate entropy}
\begin{equation}
\mathcal{S}=-\int_{\mathcal{V}_{\mathbf{x}}}   \mathrm{d}\mathbf{x} \int_{\mathcal{V}_{\mathbf{v}}}  \mathrm{d}\mathbf{v} \int_{\mathcal{V}_{\sigma}}  \mathrm{d}\sigma \ \rho\log\rho \ . \label{eq:mixing_entropy}
\end{equation}
The expression of the macrostate entropy given in Eq. (\ref{eq:mixing_entropy}) is a classical one, especially in the context of two-dimensional  turbulence \citep{miller1990,robert1991}.
A rigorous derivation of such macrostate entropy requires the use of large deviation theory; see e.g. \cite{touchette2009large} for an introduction to those tools.
A key difficulty in deriving rigorously this macrostate entropy from the usual Boltzmann entropy is that the microstates are continuous fields which contain an infinite number of degrees of freedom, and which are constrained by an infinite number of dynamical invariants.
Several discretisation procedures have been proposed to bypass this difficulty, see e.g.  \cite{michel1994large,boucher2000derivation,bouchet2010invariant,potters2013sampling,renaud2014}.
A similar formula has been derived previously by  \cite{tabak2004} in the context of non-rotating, density stratified Boussinesq fluids, in the particular case of a two-level  buoyancy configuration.
Here we have generalised this result to arbitrary buoyancy distribution, and more importantly, we have included the velocity field in the description of the microstate, which is essential to account for energy conservation.\\ 

\subsection{Computation of the most probable mascrostate, and general properties of the equilibrium states \label{sub:entro_max}}

\noindent

{ 
The first step to find the equilibrium state is to compute critical points of the variational problem given by the equilibrium theory, i.e. to find the field $\rho$ such that first variations of the macrostate entropy (\ref{eq:mixing_entropy}) around this state vanish, given the constraints of the problem given by  $\mathcal{E}[\rho]=E$,  $\mathcal{G}_{\sigma}[\rho]=G(\sigma)$,  $\mathcal{N}_{\mathbf{x}}[\rho]=1$, where $\mathcal{E}$  is the energy defined in Eq. (\ref{eq:E_rhop}),  $\mathcal{G}_{\sigma}$ is the global distribution of buoyancy defined in Eq. (\ref{eq:HHH}), and $\mathcal{N}_{\mathbf{x}}$ the local normalization of the PDF expressed in Eq. (\ref{eq:norm}). 
One needs for that purpose to introduce the Lagrange multipliers $\beta_{\mathrm{t}}$, $\gamma(\sigma)$, $\xi(\mathbf{x})$ associated with those constraints.}
Computing first variations with respect to the probability field $\rho$ yields
\begin{equation}
\delta\mathcal{S}-\beta_{\mathrm{t}}\delta\mathcal{E}+\int_{\mathcal{V}_{\sigma}} \mathrm{d}\sigma \ \gamma(\sigma)\ \delta\mathcal{G}_{\sigma}+\int_{\mathcal{V}_{\mathbf{x}}} \mathrm{d} \mathbf{x} \ \xi(\mathbf{x}) \delta\mathcal{N}_{\mathbf{x}}=0\ . \label{eq:first_var}
\end{equation}
Using the expression of the entropy, of the energy, of the global distribution of buoyancy and of the normalisation constraints given respectively in Eqs. (\ref{eq:mixing_entropy}), (\ref{eq:E_rhop}),  (\ref{eq:HHH}) and (\ref{eq:norm}), Eq. (\ref{eq:first_var}) yields  
\begin{equation}
\int_{\mathcal{V}_{\mathbf{x}}}   \mathrm{d}\mathbf{x} \int_{\mathcal{V}_{\mathbf{v}}}  \mathrm{d}\mathbf{v} \int_{\mathcal{V}_{\sigma}}  \mathrm{d}\sigma \  \left( \left(1+\log \rho\right)+\beta_t \left( \frac{\mathbf{v}^2}{2}-\sigma z\right) -\gamma\left(\sigma\right) -\xi\left(\mathbf{x}\right)\right) \delta \rho =0.\label{eq:first_varBIS}
\end{equation}
This equality is true for any $\delta \rho$, which, using the normalisation constraint in Eq. (\ref{eq:norm}),  yields the following necessary and sufficient condition for $\rho$ to be a critical point of the variational problem:
\begin{equation}
\rho\left(\mathbf{x},\sigma,\mathbf{v} \right)= \left(\frac{\beta_{\mathrm{t}}}{2\pi}\right)^{3/2} e^{-\beta_{\mathrm{t}}\frac{\mathbf{v}^2}{2} } \rho_b(z,\sigma) ,\label{eq:indep}
\end{equation}
with 
\begin{equation}
\rho_b(z,\sigma)  \equiv \frac{e^{\beta_{\mathrm{t}}\sigma z+\gamma(\sigma)}}{\mathcal{Z}(z)} ,\quad \mathcal{Z}(z)\equiv \int_{\mathcal{V}_{\sigma}}  \mathrm{d} \sigma\ e^{\beta_{\mathrm{t}}\sigma z+\gamma(\sigma)} \ . \label{eq:EQUILIB}
\end{equation}

The values of the Lagrange multipliers $\beta_{\mathrm{t}}$ and $\gamma(\sigma)$  are implicitly determined by the expression of the constraints $\mathcal{E}[\rho]=E$ and $\mathcal{G}_{\sigma}[\rho]=G(\sigma)$, given by Eq. (\ref{eq:E_rhop}) and Eq.  (\ref{eq:HHH}), respectively.

 The probability density field (\ref{eq:indep}) is expressed as a product of the probabilities  for buoyancy  and velocity, which means that $b$ and $\mathbf{u}$ are two independent quantities at equilibrium. The predicted velocity distribution is Gaussian, with zero mean ($\overline{\mathbf{u}}=0$), isotropic and homogeneous in space. It is therefore fully characterised by the local eddy kinetic energy 
\begin{equation}
e_c \equiv \frac{1}{2}\overline{\mathbf{u}^2}=\frac{3}{2}\frac{1}{\beta_{\mathrm{t}}} \label{eq:energy-beta}
\end{equation}
The inverse of $\beta_{\mathrm{t}}$ defines an effective  ``temperature'' of the turbulent field, corresponding to the turbulent agitation of fluid particles. { Remarkably, the three-dimensional nature of the flow appears only in this equation, and nowhere else. A two-dimensional  case would just have a different relation between kinetic energy and this effective temperature.}

The predicted buoyancy distribution $\rho_b$ depends only on the height coordinate $z$. The equilibrium theory predicts therefore that the local fluctuations of buoyancy are invariant in the horizontal. {  It means that in the remaining of this paper, the quantities $\overline{\cdot}$ can be interpreted either as a local coarse-graining or as an horizontal average. Similarly, the quantity $\rho_b$ can be interpreted either as a local distribution of buoyancy or as the distribution of buoyancy over an horizontal plane.}

Eq. (\ref{eq:EQUILIB}) relates the mean buoyancy profile and its fluctuations to the effective turbulent temperature. 
Buoyancy moments are defined at  each height  in terms of $\rho_b(z,t)$ as
\begin{equation}
\overline{b^n}(\mathbf{x}) \equiv \int_{\mathcal{V}_{\sigma}}  \mathrm{d}\sigma\  \sigma^n \rho_b \ .  \label{eq:bar_b} \label{eq:eq_statmom}
\end{equation} 
From Eq. (\ref{eq:EQUILIB})  we get the relations 
\begin{equation}
\overline{b}=\frac{1}{ \beta_{\mathrm{t}}}\frac{\mathrm{d} \log \mathcal{Z}}{\mathrm{d} z} \ , \quad  \overline{b^2}-\overline{b}^2=\frac{1}{ \beta_{\mathrm{t}}^2} \frac{\mathrm{d}^2 \log \mathcal{Z}}{\mathrm{d} z  ^2} \ . \label{eq:b_b2_Z}
\end{equation} 
Using those expressions and Eq. (\ref{eq:energy-beta}), one gets finally an expression relating the mean buoyancy profile to the ratio of the buoyancy fluctuations to the kinetic energy fluctuations:
\begin{equation}
\frac{\mathrm{d} \overline{b}}{\mathrm{d} z}={3}\frac{\overline{b^{2}}-\overline{b}^{2}}{2e_c} \ .  \label{eq:beta_fluct}
\end{equation}
In the case of a strong stratification, the local variance of buoyancy is proportional to the small vertical displacement of fluid elements, so this relation can be interpreted as an equipartition between kinetic and potential energy fluctuations, as further discussed in section \ref{sec:mixing_efficiency}.

{The equilibrium state has a peculiar spatial structure: the buoyancy field $b$ is characterised by a smooth coarse-grained buoyancy profile $\overline{b}(z)$ superimposed with small-scale buoyancy fluctuations. More precisely, the theory predicts that when performing a local coarse-graining of the microscopic buoyancy and velocity fields at a scale $l$ (the scale of the macrocell within the framework of our discrete model depicted in fig. 1), the small-scale fluctuations are confined at scales smaller than the coarse-graining scale $l$, no matter how small the coarse-graining length scale $l$.}

{  In the case of a decaying experiment with weak molecular viscosity and diffusivity, the subgrid-scale velocity fluctuations of the equilibrium state correspond to the amount of kinetic energy that will be dissipated by viscosity during the whole decay, the subgrid-scale buoyancy fluctuations of the equilibrium state correspond to the amount of buoyancy fluctuations locally dissipated by diffusivity during the whole decay. As a result, the equilibrium state $\overline{b}$ corresponds to the background buoyancy profile that will be measured after a mixing event, once the system has reached a state of rest in a decaying experiment. The underlying hypothesis is that the system reaches the equilibrium states before molecular effects become important.
}  


\section{Computation of mean equilibrium buoyancy profiles}

\subsection{The two-level case}\label{sub:twolevels}

\noindent

We discussed in the previous subsection the general case with a continuum of buoyancy levels. 
In the particular case with a finite number of buoyancy levels (say $K$ levels $\sigma_k$ with $1\le k\le K$), the buoyancy field is described at a macroscopic level by $p_k(\mathbf{x})$, which is the probability of measuring the level $\sigma_k$ at point $\mathbf{x}$ with $\sum_{k=1}^K p_k(\mathbf{x})=1$, see Appendix \ref{app:A}.
The same arguments as in subsection \ref{sub:entro_max} for the computation of the equilibrium state then yields 
\begin{equation}
p_k(z)  \equiv \frac{e^{\beta_{\mathrm{t}}\sigma_k z+\gamma_k}}{\sum_{k=1}^{K} e^{\beta_{\mathrm{t}}\sigma_k z+\gamma_k} }\ ,  \quad \beta_t=\frac{3}{2e_c} , \label{eq:EQUILIB_discret}
\end{equation}
where the values of the Lagrange multipliers $\beta_{\mathrm{t}}$ and $\{\gamma_k\}_{1\le k\le K}$  are implicitly determined by the energy constraint and conservation of the total volume occupied by each buoyancy  level $\sigma_k$. 

Let us restrict ourselves to the case of an initial state composed of two buoyancy levels in equal proportion with
\begin{equation}
\forall \mathbf{x} \in \mathcal{V}_{\mathbf{x}},\ b(\mathbf{x}) \ \in \ \left\{ -\frac{\Delta b}{2},\ \frac{\Delta b}{2}\right\} . \label{eq:def_2lev}
\end{equation}
The only dimensionless parameter of the problem within the statistical mechanics framework is given by the global Richardson number based on the total height $2H$, buoyancy jump $\Delta b$ and square of velocity fluctuations $2e_c$:
\begin{equation}
Ri \equiv \frac{H \Delta b}{e_c} \ .\label{eq:def_Ri}
\end{equation}
{ This global  Richardson number based on the domain height $H$  is different from the bulk Richardson number $Ri_{b}=\Delta b L_t/ e_c=(L_t/H) Ri $ based on the turbulent length scale $L_t$, which is commonly  used  in the context of turbulent mixing in stratified fluids; see e.g. \citep{fernando91}. The statistical mechanics prediction depends only on the total energy, not on its injection scale  $L_t$. This point will be further discussed in section 4.4.}

We denote $p_+(z)$ the probability of measuring $\Delta b /2$ at height $z$.
According to the notation used in Eq. (\ref{eq:EQUILIB_discret}), we get $\sigma_1=-\Delta b/2$, $\sigma_2=\Delta b/2$, $p_1=1-p_+$, $p_2=p_+$, with 
\begin{equation}
p_+(z)=\frac{e^{\frac{3Ri}{4} \frac{z}{H} }}{e^{-\frac{3Ri}{4}  \frac{z}{H} }+e^{\frac{3Ri}{4} \frac{z}{H}  }}  \ ,\label{eq:pz}
\end{equation}
{ where we have used the symmetry with respect to $z=0$  ($p+(z)=-p(-z)$) and the fact that the two buoyancy  levels are in equal proportions ($\int_{-H}^{+H} \mathrm{d} z p_+=\int_{-H}^{+H} \mathrm{d} z p_-$) to eliminate the Lagrange parameters $\gamma_1,\gamma_2$ in Eq. (\ref{eq:EQUILIB_discret}).}

Equation (\ref{eq:pz}) is reminiscent of the Fermi-Dirac distribution. 
{  Indeed, the conservation of buoyancy plays here the same role as the exclusion principle for the statistics of fermions: within the framework of the discretised model depicted in Fig. \ref{fig:schema}, the buoyancy carried by a fluid particle at a given grid point can only take one value among  $-\Delta b/2$ and $\Delta b/2$. }
Following this analogy, the buoyancy  field is a collection of fluid particles carrying the potential energy $e_p=\pm  \sfrac{1}{2} z \Delta b $, with a Fermi level $\varepsilon_f=0$, in thermal contact with a heat bath characterised by the inverse temperature $\beta_{\mathrm{t}}$.

Using Eq. (\ref{eq:pz}) and  Eq.  (\ref{eq:energy-beta}), the mean density profile $\overline{b}= \frac{\Delta b}{2} p_+-\frac{\Delta b}{2}(1-p_+)$ is expressed as
\begin{equation}
\overline{b}(z) = \frac{\Delta b}{2}\tanh\left(\frac{3 Ri}{4}\frac{z}{H} \right)\ .  
\label{eq:mean_reduced_density}
\end{equation}
Large global  Richardson numbers $Ri \gg 1$ correspond to sharp interfaces: the kinetic energy is too small to allow for large excursion of fluid particles away from the rest position.  
By contrast, small global Richardson numbers $Ri\ll 1$ correspond to a homogenised buoyancy field: the total kinetic energy is much larger than the energy required to mix the buoyancy field.
This tanh profile was  previously obtained by \cite{tabak2004} using similar arguments, but without relating the effective temperature to the kinetic energy of the flow in a consistent theory. Our approach allows for a direct interpretation of the effective temperature of the flow as the local turbulent kinetic energy, which will make possible quantitive estimate for mixing efficiency.

\subsection{A relaxation equation towards the equilibrium states}

\noindent

The expression for the equilibrium state given in Eq. (\ref{eq:EQUILIB})  requires the knowledge of the Lagrange multipliers $\gamma(\sigma)$ and  $e_c$, which depend implicitly on the constraints $G(\sigma)$ and $E$. 
This makes analytical computations of those equilibria very challenging. 
Solutions may be obtained in particular cases, such as for the two-level configuration analysed in subsection \ref{sub:twolevels}, but more generally it must be determined numerically.%

We devise for that purpose an algorithm based on a \emph{maximum entropy production principle}, which was introduced by  \cite{robert1992relaxation} in order to compute equilibrium states of two-dimensional Euler flows.
The idea of the algorithm is to consider a time dependent probability distribution function 
\begin{equation}
\rho\left(\sigma,\mathbf{x},\mathbf{v},t\right)= \left(\frac{3}{4 \pi e_c(t)}\right)^{3/2} e^{-\frac{3}{2 e_c(t)} \frac{\mathbf{v}^2}{2} } \rho_b(z,\sigma,t) ,\label{eq:ansatz}
\end{equation}
where the pdf $\rho_b(z,\sigma,t)$ and the local kinetic energy $e_c(t)$ depend on time, and can be different from the pdf and the kinetic energy of the actual equilibrium state. 
We derive in Appendix \ref{app:B} a dynamical equation for $\rho_b$ that conserves the total energy and the global distribution of buoyancy levels, while maximising the entropy production at each time:
\begin{equation}
\partial_{t}\rho_b=\partial_{z}\left[ D \left(\partial_{z}\rho_b-\frac{3}{2e_c}\left(\sigma-\overline{b}\right)\rho_b\right)\right]\ ,\label{eq:forme_relaxation}
\end{equation}
where $D$ is an arbitrary positive diffusion coefficient. The kinetic energy $e_c$ defined in Eq. (\ref{eq:energy-beta}) is expressed in terms of the total energy $E$ and the buoyancy profile $\overline{b}(z,t)$ by using Eq. (\ref{eq:E_rhop}):
\begin{equation}
e_c=\frac{E}{V}+\frac{1}{2H}\int_{-H}^{+H} \mathrm{d} z \ \left(\overline{b} -b_s \right)z, \label{eq:ec_E_Bbar}
\end{equation}
with $V$ the volume of the flow domain.

Maximising the entropy production ensures that the system relaxes towards an equilibrium state.
Indeed, using Eq. (\ref{eq:EQUILIB})-(\ref{eq:energy-beta}) and the first equality in Eq. (\ref{eq:b_b2_Z}), the equilibrium states can be written as
\begin{equation}
\rho_b(\sigma,z)=\rho_b(\sigma, 0)  e^{\frac{3}{2e_c}\left(\sigma z -\int_0^z \mathrm{d} z'\ \overline{b}(z') \right)} \ , \label{eq:eq_stat} 
\end{equation}
which is also the expression of any stationary solution of Eq. (\ref{eq:forme_relaxation}).
According to equation (\ref{eq:forme_relaxation}) the equilibrium state can be interpreted as the result of a compensation between usual turbulent diffusion and a drift term corresponding to restratification of buoyancy fluctuations.
We stress that the convergence towards equilibrium depends on the parameter $D$, but that the equilibrium itself does not depend on this parameter. This is why is can be chosen arbitrarily. 

\begin{figure}
\begin{center}
\includegraphics[width=\textwidth]{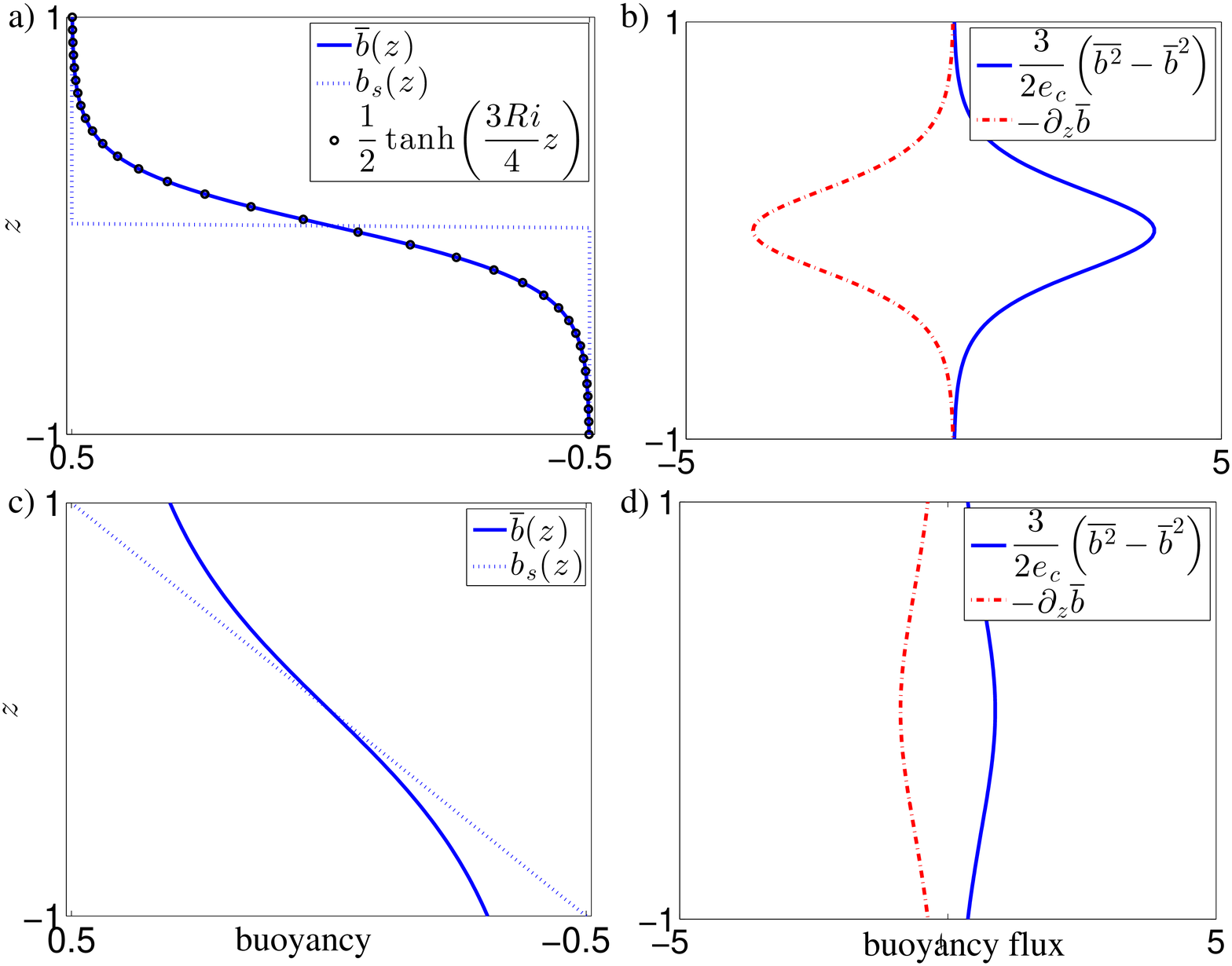}
\end{center}
\caption{a) Plain blue line: equilibrium state $\overline{b}(z)$ computed numerically in the case $Ri=10$, where $Ri=H\Delta b/e_c$ is the global Richardson number. Here $H=1$, $\Delta b=1$. The dotted blue line: corresponding background buoyancy profile $b_s(z)$ (here a two-layer case). Black circles : analytical expression from Eq. (\ref{eq:mean_reduced_density}) for the equilibrium state of the two-level system. The buoyancy increases from right to left on the horizontal axis. b) Compensation of the downgradient buoyancy flux with the restratification term proportional to buoyancy fluctuations (with $D=1$). The total buoyancy is the sum of those two terms, which is zero at equilibrium. d)  Same as a,b in the case of an initial linear background buoyancy profile (no analytical predictions in that case).
 \label{fig:rho_eq_exple}}
\end{figure}

Assuming that the initial energy  $E$ injected into the system and that the background buoyancy profile $b_s(z)$ are known, one can then use the relaxation algorithm  (\ref{eq:forme_relaxation}), starting from the state  
\begin{equation}
\rho_b(z,\sigma,0)=\delta(b_s(z)-\sigma),\quad e_c(0)=\frac{E}{V} .\label{eq:initial_state}
\end{equation}
Equation (\ref{eq:forme_relaxation}) is an integro-differential equation, because the local kinetic energy is a functional of the macroscopic vertical buoyancy profile.
Its numerical implementation is much easier assuming that $e_c$ is a constant.
One then loses energy conservation, but the equation still conserves the global buoyancy distribution, assuming no buoyancy fluxes at the upper and lower boundaries. It can be shown that this process minimises the free-energy production defined as $\dot \mathcal{F}=-\dot \mathcal{S}+\beta_t\dot \mathcal{E}$, where the upper dot stands for a time derivative, and where $\beta_t={3}/({2e_c})$ can be interpreted as the inverse of an effective turbulent temperature.
Indeed, assuming constant local kinetic energy amounts to a computation of the equilibrium state within the canonical ensemble where the "heat bath" is provided by turbulent agitation.
{  In order to solve numerically Eq. (\ref{eq:initial_state}) with constant $e_c$, we first  assume a discretisation of the global buoyancy distribution into $N_{\sigma}$ buoyancy levels denoted $\sigma_n$ with $1 \le n\le \ N_{\sigma}$. Denoting $\rho_{b,n}(z,t)$ the probability to measure the level $\sigma_n$ in the vicinity of  height $z$ at time $t$, we obtain a system of one dimensional parabolic partial differential equations for $\left\{\rho_{b,n}(z,t)\right\}_{1 \le n\le \ N_{\sigma}}$, which can be solved using standard numerical procedures.
This dynamical system is integrated in time until a steady state is reached. This steady state is the equilibrium state. 
Once the equilibrium state associated with a given value of $e_c$ is computed, it is straightforward to compute its total energy $E$ using Eq. (\ref{eq:E_rhop}). One can then check that varying $e_c$ from $0$ to $+\infty$ amounts to varying $E$ from $0$ to $+\infty$. This procedure therefore provides the complete set of equilibria associated with any given background buoyancy profile.}\\ 
We show in Fig. \ref{fig:rho_eq_exple} two examples of equilibrium states computed by this procedure, assuming no buoyancy fluxes at the upper and lower boundaries. 
Panels a,b corresponds to the two-level configuration. As expected from Eq. (\ref{eq:mean_reduced_density}), the mean equilibrium buoyancy profile is characterised by a tanh shape in that case. Panel b confirms that this equilibrium state may be interpreted as the result of a balance between a classical downgradient term $-D\partial_z {\overline{b}}$ modelling turbulent transport and a term $D (3/2e_c) \left(\overline{b^2}-\overline{b}^2\right)$ modelling restratification. %

Panels c,d correspond to the more complicated case of a linear profile for the background buoyancy profile, for which no analytical results exist. Just as in the two-layer case, we see enhanced buoyancy fluctuations in the domain bulk. This numerical method can easily be applied to any background buoyancy profile, and will be applied in next section to the computation of mixing efficiency.

\section{Computation of mixing efficiency in decaying flows\label{sec:mixing_efficiency}}

{ 

\subsection{Irreversibility and mixing efficiency \label{sub:irr}}

\noindent

We argue in the following that the computation of the equilibrium states for the inviscid, adiabatic system can be used to obtain quantitive predictions for the efficiency of mixing in decaying stratified turbulence. %

The first assumption is that molecular viscosity and diffusivity only play a secondary role  in the limit of large Reynolds and P\'eclet numbers. More precisely, we assume that the time scale to reach the equilibrium state of the inviscid, adiabatic dynamics is smaller than the typical time scale of dissipative effects. In other words, inertial dynamics govern the amount of small-scale velocity and buoyancy  fluctuations that are created on a short time scale, and  the only effect of viscosity and diffusivity is to smooth-out  these  fluctuations on a longer time scale. 

The second assumption is that the flow system evenly explores phase space through turbulent stirring, which is necessary to use statistical mechanics predictions.  According to the theory,  the macroscopic buoyancy profile $\overline{b}$ and the local distribution of small-scale fluctuations do not evolve in time anymore once the equilibrium state is reached: the equilibrium state is an attractor for the dynamics. In that respect, the purely inertial, inviscid and adiabatic dynamics is irreversible.  In other words, even if the process described by the equilibrium theory is pure stirring, it implies irreversible mixing of the buoyancy field at a coarse-grained level. 
Assuming that this stationary property of $\overline{b}$ persists in the presence of weak viscosity and weak dissipation, we see from Eq. (\ref{eq:beta_fluct}) that the rate of local small-scale kinetic energy dissipation $\mathrm{d} \log e_c/\mathrm{d} t$  should be equal to the rate of dissipation for  the local variance of local buoyancy fluctuations $\mathrm{d} \log \left(\overline{b^2}-\overline{b}^2\right)/\mathrm{d} t$.

Let us assume that a given amount of energy denoted $E_{inj}$ is injected in a fluid initially at rest, characterised by  a background buoyancy profile  $b_{s}(z)$. The injected energy may either be purely kinetic (through mechanical stirring) or purely potential (for instance by turning the tank upside down into an unstable configuration).  Once the equilibrium state is reached, part of this energy is carried by small-scale velocity fluctuations, and the remaining part is used to maintain the potential energy of the system at a higher value than the potential energy of the background state. The transfer of of part of the initial energy present at a coarse-grained level into subgrid-scale  (fine-grained) fluctuations is very much similar to the effect of viscosity, which transfers energy from the degrees of freedom of the fluid motion to these of thermal fluctuations.
The total kinetic energy carried by the equilibrium state is  denoted $E_c=V e_c$  with $e_c$ the local kinetic energy  density, homogeneous in space. This kinetic energy takes the form of small-scale fluctuations, that will be eventually dissipated  in a decaying experiment with weak viscosity,  and the quantity $E_c$ can then be interpreted as the temporal integral of viscous dissipation.

Turbulent stirring implies rearrangements of fluid parcels, and such rearrangements from $b_s(z)$ to $b(x,y,z)$ are necessarily associated with an increase of potential energy
\begin{equation}
E_{p} = -\int_{\mathcal{V}_{\mathbf{x}}}  \mathrm{d} \mathbf{x}   \ \left( b -b_{s} \right) z .
\label{eq:ape_def} 
\end{equation}
At equilibrium, this quantity can be expressed in terms of the macroscopic buoyancy profile $\overline{b}$ which depends only on $z$:
\begin{equation}
E_{p}= -\frac{V}{2H} \int_{-H}^{+H} \mathrm{d} z \ \left(\overline{b} -b_s\right)z \  .\label{eq:def_Eape}
\end{equation}
This definition is equivalent to the classical definition of the available potential energy. However, as explained above, the convergence towards the equilibrium buoyancy profile is irreversible. Once the equilibrium is reached, the available potential energy $E_p$ has been irreversibly transferred to smaller scales, and can not be transferred anymore into another form of energy. It would inescapably result into molecular mixing in the presence of molecular diffusion. In that case, $E_p$ would corresponds to the increase of the background potential energy, which is consistent with \cite{winters1995available}.

We define the mixing efficiency as
\begin{equation}
\eta \equiv \frac{E_{p}}{E_p+E_c} ,\label{eq:eta2}
\end{equation}
where $E_p+E_c=E_{inj}$ is the total energy injected into the system. This definition of mixing efficiency is bounded between $0$ and $1$. Since $E_c$ is the total amount of kinetic energy lost at small scale, and since $E_p$ corresponds to an irreversible increase of potential energy according to statistical mechanics theory, our definition of $\eta$  is equivalent to the long time limit of the \textit{cumulative mixing efficiency}  \citep{peltier2003}, or to the \textit{integrated flux Richardson number} \citep{linden1979}.

{  We stress finally that the equilibrium theory does not predict a temporal evolution for the system but just the final outcome of turbulent stirring under the assumption of random evolution without forcing and dissipation. It provides therefore a global  (integrated over the whole domain) and cumulative (integrated over sufficiently large time)  prediction for the efficiency of mixing.} 

\subsection{Numerical computation in the general case}

\noindent
{  
We show in Fig. \ref{fig:mixing_efficiency} how the mixing  efficiency $\eta$  varies  with the global Richardson number $Ri=H\Delta b/e_c$, with $\Delta b=b_s(H)-b_s(-H)$. We consider two different buoyancy profiles $b_s(z)$: case (a) is the two-level configuration corresponding to a background profile with two homogeneous layers of equal depth, for which an analytical solution exits; case (b) corresponds to  a  linear background buoyancy profile.
Considering those two cases allows us to show very different behaviour for the  variations of  mixing efficiency as a function of the Richardson number $Ri$. 
The kinetic energy $e_c$ appearing in the Richardson number is not a control parameter, but one can check \textit{a posteriori} that $E_c=V e_c$ is always of the same order of magnitude as the injected energy $E_{inj}$, which is a control parameter. In a direct numerical simulations with non-zero viscosity, $E_c$ would be the actual amount  of kinetic energy  dissipated during the turbulent decay. 

We see in Fig. \ref{fig:mixing_efficiency} that whatever the background buoyancy profile, the equilibrium buoyancy profile $\overline{b}$ can be considered as almost completely  homogenised in the low Richardson number 
limit $Ri\ll 1$. In that case, most of the injected energy is lost in small-scale velocity fluctuations with $E_c=  E_{inj} $ and the fluid is well mixed, so that $\overline{b}$ is a constant,  and Eq. (\ref{eq:def_Eape}) reduces to $E_{p}=   \frac{V}{2H} \int_{-H}^{+H}  b_s z \mathrm{d} z  $. The mixing efficiency is then given by  
\begin{equation}
\eta  =_{Ri\ll 1} {Ri} \ \Xi[b_s] \quad  \text{with}\quad  \Xi[b_s]\equiv\frac{1}{2\Delta bH^2}\int_{-H}^{+H}  b_s z \mathrm{d} z \ \quad \text{when } Ri\ll 1 . 
\end{equation}
The numerical coefficient $\Xi[b_s]$ is bounded in $[0\ 1]$ and characteristic of the shape of the background buoyancy profile, hence of the distribution of available densities. It is equal to $0$ for a homogeneous fluid, $1/6$ for a linear stratification and $1/4$ for a two-layer system. %
Whatever this background buoyancy profile, the mixing efficiency scales linearly with the Richardson number is the limit of weak Richardson numbers.

By contrast, we see in Fig. \ref{fig:mixing_efficiency} that the large Richardson behaviour of the mixing efficiency depends drastically on the background buoyancy profile $b_s$: the mixing efficiency decreases to zero with increasing Richardson numbers in the two-level case of Fig. \ref{fig:mixing_efficiency}a, while it increases to an asymptotic value close to $0.25$ in the linearly stratified case of  Fig. \ref{fig:mixing_efficiency}b. 
We show analytically in the next subsection that an asymptotic value of $\eta=0.25$ is indeed expected in a low energy limit, as a consequence of energy equipartition, provided that the stratification of the background profile is always strictly positive ($\partial_z b_s >0$ for $-H\le z \le H$).  

\begin{figure}
\begin{center}
\includegraphics[width=\textwidth]{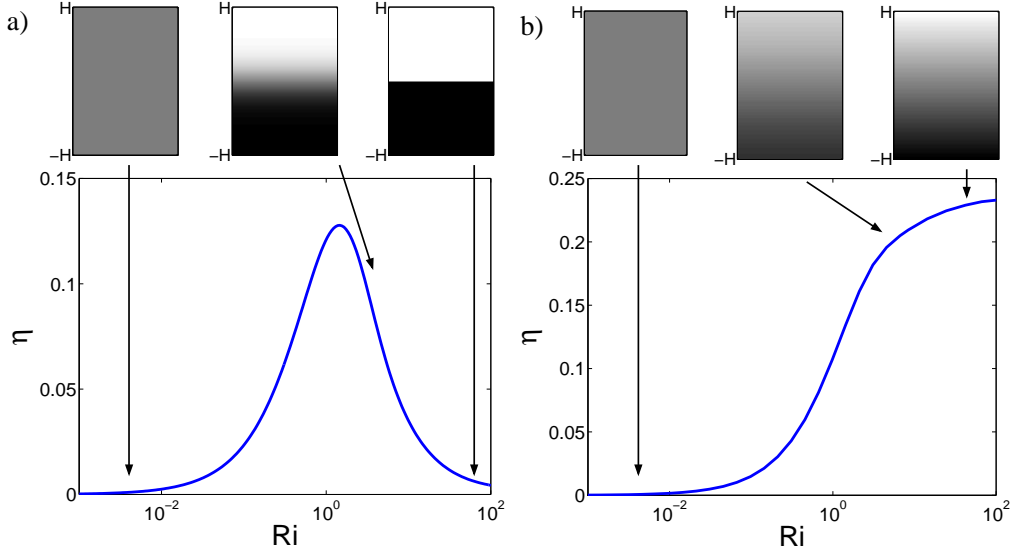}
\end{center}
\caption{Variation of the mixing efficiency $\eta=E_{p}/E_{inj}$ with  the Richardson number $Ri=H\Delta b/e_c$  (a) for a background buoyancy profile with two homogeneous layers, (b)  for an initial linear background buoyancy profile. The three insets show the equilibrium buoyancy field  $\overline{b}$ for $Ri=0.07, 7, 70$. \label{fig:mixing_efficiency}}
\end{figure}

\subsection{Energy equipartition and mixing efficiency for high Richardson numbers \label{sec:mixing_efficiency}}

\noindent

The potential energy  $E_p$ defined in Eq. (\ref{eq:ape_def}) is a linear functional of $b-b_s$, which is \textit{a priori} sign indefinite. However, the conservation of the global distribution of buoyancy levels (prescribed by $b_s$) provides a strong constraint on admissible buoyancy levels $b$, and hence on admissible values for $E_p$. A direct consequence of these conservation laws it that the potential energy is strictly positive unless $b=b_s$.  Denoting $Z_s(b_s)$ the height of fluid particles carrying buoyancy level $b_s$ in the background buoyancy profile,  using an asymptotic expansion in terms of $b-b_s$ and assuming $\rm{d}b_s/\rm{d}z=b_s^{\prime } > 0$,  one can use the conservation laws related to buoyancy (Casimir functionals) to obtain an explicit quadratic form for the potential energy  in a weak energy limit \citep{shepherd1993unified}: 
\begin{equation}
E_{p} =  \frac{1}{2} \int_{\mathcal{V}_{\mathbf{x}}} \mathrm{d} \mathbf{x}  \frac{\left(b-b_{s}  \right)^2}{ b_{s}^{\prime}} +O\left( Z_s^{\prime\prime} \left(b-b_s\right)^3 \right)\ .  \label{eq:ape_appbis0}
\end{equation}
The quadratic part is also the classical expression of the potential energy for internal gravity waves, derived for instance in  \citep{GillBook}. 

Decomposing the spatial integral of Eq. (\ref{eq:ape_appbis0})  into a sum of integrals over each macrocell of the discrete model depicted in Fig. 1 and taking the limit of an infinite number of macrocells, the potential energy can be expressed in terms of the local variance of buoyancy fluctuations:
\begin{equation}
E_{p} =\frac{V}{4H}  \int_{-H}^{+H} \mathrm{d} z\  \frac{\overline{b^2}-\overline{b}^2}{ b_{s}^{\prime}} +O\left( Z_s^{\prime\prime} \overline{\left(b-b_s\right)^3} \right)\ ,    \label{eq:ape_appbis}
\end{equation}
The variance of buoyancy fluctuations $\overline{b^2}-\overline{b}^2$ is related to the local kinetic energy $e_c$ and the local buoyancy gradient $\rm{d} \overline{b}/\rm{d}z$ though Eq. (\ref{eq:beta_fluct}). Inserting this equation into Eq. (\ref{eq:ape_appbis}), using $\rm{d} \overline{b}/\rm{d}z = { b_{s}^{\prime}}$ (which is valid for sufficiently large Richardson numbers),  and $e_c=E_c/V$:
\begin{equation}
E_{p} =  \frac{E_{c}}{3}  , \label{eq:relate_ape_ec}
\end{equation}
which shows equipartition of the energy between the available potential energy and the three degrees of freedom of the kinetic energy. A direct consequence of energy equipartition is thus 
\begin{equation}
\eta  =  \frac{1}{4} . \label{eq:eta_one_quarter}
\end{equation}
This result is a direct consequence of the quadratic form of the energy obtained in Eq. (\ref{eq:ape_appbis}), which  relies on the assumptions (i) that $b_s^{\prime}$ is strictly positive and bounded (ii) that $Z_s^{\prime \prime}$ is bounded (iii)  that $b$ remains sufficiently close to $b_s$.
  
Importantly, the hypotheses (i) and (ii) are not satisfied when the background buoyancy profile contains homogeneous layers of fluids, as for instance in the case depicted in Fig. \ref{fig:mixing_efficiency}a.  In order to evaluate when the assumption (iii) is valid, one can estimate the typical value of  $b-b_s$ at a given point  as the root mean square of local buoyancy fluctuations $\left(\overline{b^2}-\overline{b}^2\right)^{1/2}$. Using Eq. (\ref{eq:beta_fluct}),  $\partial_z \overline{b}= \partial_z b_{s}$  and $Ri\sim  H^2 b_s^{\prime} /e_c$ yields then $\left(b-b_s \right) \sim Ri^{-1/2}$. 

We conclude that $\eta=0.25$ is expected in the limit of large  Richardson number, when the background buoyancy profile  is strictly increasing with height.

\subsection{Comparison with previous studies of the efficiency of mixing \label{sub:discussion}}

\noindent

Despite the large number of numerical and experimental studies devoted to the understanding of mixing efficiency, there are only few theoretical results yielding predictions for the variations of mixing efficiency with the Richardson number. In the context of shear-stratified turbulence, dimensional analysis was used by \cite{townsend1958effects} to model the variation of mixing efficiency with the gradient Richardson number, and upper bounds for the mixing efficiency have been derived rigorously by \cite{caulfield2001maximal}.

A mixing efficiency efficiency $\eta=0.25 $ was obtained by a phenomenological model due to  \cite{mcewan1983kinematics}, based on purely kinematic arguments. Those predictions were found to be consistent with experimental observations of mixing efficiency following an internal wave-breaking event  \citep{mcewan1983internal}.  The argument is the following: take a continuously stratified fluid at rest, and exchange two particle fluids $a$  and $b$ of volume $\delta V$ with buoyancy difference $\Delta b=b_b-b_a$ and height difference $\Delta z=z_b-z_a$, with $\delta V/ V \ll \Delta z/H$. Then consider the small displacement limit $\Delta z \rightarrow 0$, which, as explained in previous sections, corresponds to a weak energy limit, or equivalently to a large Richardson number limit, for which $\Delta b =   \Delta z \rm{d} \overline{b}/\rm{d}z $. Given that the injected energy is under the form of available potential energy only, the initial kinetic energy is zero, with $ E_{inj} =   b_s^{\prime} (\Delta z)^2 \delta V$.  \cite{mcewan1983kinematics} then argued that the two displaced fluid particles will be stirred and mix together until homogenisation of their buoyancy, and that the two fluid particles carrying buoyancy $(b_a+b_b )/2$ will "sediment" to their rest position $z= (z_a+z_b)/2$. The available potential energy of the final state is then $E_{p} =  b_s^{\prime} (\Delta z/2)^2\delta V$, which corresponds to mixing efficiency $\eta= 0.25$.  

Strikingly, several numerical studies have also reported convergence of mixing efficiency  towards  $\eta= 0.25$ at large Richardson numbers; see e.g. \cite{maffioli2016mixing,Venayagamoorthy2016} and references therein\footnote{\cite{maffioli2016mixing} report a mixing coefficient $\Gamma=\eta/(1-\eta)=0.33$ in the limit of small Froude numbers, which corresponds to $\eta=0.25$ in the limit of large Richardson numbers.}. It is remarkable that the statistical mechanics theory in the large Richardson number limit also yields $\eta=0.25$, provided that the initial buoyancy is strictly monotonic. We stress that the only assumption underlying the equilibrium theory is that the system evenly explores the phase space: there is neither dynamics nor kinematics involved in the derivation of this result. By contrast, the approach of \cite{mcewan1983kinematics} relies on the choice of a peculiar kinematic model. 

\cite{mcewan1983kinematics} also discussed the case of two homogeneous layers separated by a linear pycnocline of thickness $\delta$. He found that mixing efficiency vanishes when considering first the limit $\delta \rightarrow 0$ and second the limit of  Large Richardson numbers $Ri \rightarrow + \infty$. This is again fully consistent with the statistical mechanics predictions for the mixing efficiency in the two-level case depicted in Fig. \ref{fig:mixing_efficiency}a. Indeed, it corresponds to the case of a background buoyancy profile with an infinity sharp interface. The mixing efficiency vanishes in the limit of infinite Richardson numbers because  kinetic energy is spread equally over the whole domain at equilibrium, while  buoyancy mixing is confined to a thin layer surrounding the buoyancy interface, with a thickness that decrease with the Richardson number, as explained in subsection \ref{sub:twolevels}. 

We stress that  the statistical theory makes possible predictions for global, cumulative mixing efficiency in decaying turbulence whatever the Richardson number, and whatever the background buoyancy profile. In particular, it predicts a bell shape for $\eta(Ri)$ in the two-layer case, with a maximum $\eta=0.15$, and a monotonic increase of $\eta(Ri)$ in the linear case from $\eta=0$ to $\eta=0.25$, as shown in Fig. \ref{fig:mixing_efficiency}. The bell shape for $\eta(Ri)$ has been reported in decaying experiments performed by dropping a grid in a two-layer stratified fluid  \citep{linden1980mixing}, but how to estimate the amount of energy injected into the system in experiments continue to be debated, see e.g. \cite{huq1995turbulence}. The monotonic increase of cumulative mixing efficiency in the case of a linear background buoyancy profile seems a robust result in laboratory and numerical experiments, see e.g. \cite{stretch2010mixing}. However, the equilibrium theory does not account for layering which is often observed in the strongly stratified regime $Ri\gg 1$ \citep{rehmann2004mean}. In any case, the statistical mechanics prediction that mixing efficiency depends strongly on the global shape of the background buoyancy profile, and not only on the local buoyancy gradient is consistent with observations by \cite{holford1999turbulent}.

There is however one result that does not depend on the shape of the buoyancy profile: according to the equilibrium theory, the mixing efficiency should increase linearly with the  Richardson numbers in the limit of weak Richardson numbers. This scaling law can be simply understood as a consequence of the fact that buoyancy behaves as a passive tracer in this limit \citep{holford1999turbulent}. This linear scaling has been also reported by \cite{maffioli2016mixing} in forced-dissipative numerical experiments, who also provide complementary arguments based on cascade phenomenology.

{  The observations mentioned in the above paragraphs provide  support for the statistical mechanics predictions. Other observations however point to limitations of the theory.}  For instance, the predicted value of mixing efficiency in decaying turbulence depends on the total energy injected into the system, but not on how the energy is injected. Yet different values of mixing efficiency have been reported in laboratory and numerical experiments performed with different energy injection mechanism. A value $\eta \approx 0.2$ was reported in decaying  sheared-stratified fluids with a Richardson number of order one \citep{peltier2003}. This value is somewhat  larger than the cumulative mixing efficiency  $\eta= 0.11$ observed in  lock-exchange experiments \cite{prastowo2008mixing,ilicak2014energetics}, and smaller than the cumulative mixing efficiency $\eta \approx 0.5$  reported in the framework of Rayleigh-Taylor experiments \citep{dalziel2008mixing,wykes2014efficient}. Importantly, these different values for mixing efficiency  do not depend only on the Richardson, Reynolds and P\'eclet numbers. This suggests that the mechanism of injection plays an important role.  
 
A heuristic way to discuss more precisely the role of the injection mechanism  in relation with the ergodicity hypothesis  is to consider the parameter $L_t /H $, i.e.  the  ratio  of  the energy injection length scale  to the domain scale. In the context of two-dimensional turbulence, this parameter has been proven useful to discuss the relevance of the ergodic hypothesis underlying statistical mechanics theory \citep{Pomeau94,tabeling2002two,venaille2015violent}. Denoting $T_{tran}$ the typical time scale to move a fluid particle from the top to the bottom of the tank through turbulent transport, and calling $T_{diss}=L_t/U$ the typical  time scale for the dissipation rates for local buoyancy fluctuations through direct turbulent cascade, the system can  explores the phase  space only if $T_{trans}<T_{diss}$. Modelling turbulent transport as an effective eddy viscosity or eddy diffusivity $UL_t$ yields $T_{trans}=H^2/(UL_t)$  and then the necessary condition $H<L_t$ for ergodicity. Given that $L_t$ can not be larger than the domain height, we see that the condition for sufficient mixing in phase space will only be marginally satisfied when $H=L_t$, and will not be satisfied when $L_t \ll H$. }

Finally, the equilibrium theory applies in principle to flow systems in the limit of infinitely large Reynolds and P\'eclet numbers. Even if it is natural to expect that the dissipation rate of buoyancy and kinetic energy become independent from the value of molecular viscosity and diffusivity when they are sufficiently weak, numerical and laboratory experiments are often performed in intermediate regimes for which those parameters may influence the mixing efficiency, see e.g.  \cite{shih2005parameterization,lozovatsky2013mixing,bouffard2013diapycnal,salehipour2015diapycnal}.}

 \section{Conclusion}

{  

\noindent

We have addressed the problem of mixing efficiency from the point of view of equilibrium statistical mechanics. The theory predicts that the unforced, inviscid, adiabatic dynamics is attracted towards a state characterised by small-scale velocity fluctuations carrying kinetic energy, and by a  smooth, monotonic buoyancy profile upon which are small-scale buoyancy fluctuations. Although the whole dynamics is adiabatic, the buoyancy field is irreversibly mixed at a coarse-grained level, no matter how small the coarse-grained scale. In addition, the coarse-grained fields predicted by the theory are stationary and characterised by a stable buoyancy profile. The theory also predicts velocity fluctuations to be Gaussian, isotropic, homogeneous in space, and that the buoyancy fluctuations are homogeneous on horizontal planes. 

The input of the theory is the total energy injected initially into the system, and the global distribution of buoyancy levels, or equivalently the background buoyancy profile. The output of the theory is the probability to measure a given buoyancy level at each height. We provide explicit computations of the equilibria in limiting cases, and implement an algorithm based on a maximum entropy production which determines the equilibrium state for any background buoyancy profile. This allows us to  compute a cumulative mixing efficiency defined as the ratio of the potential energy gained by the system to the total energy injected into the system. Importantly, the potential energy effectively gained by the system is the potential energy of the coarse-grained buoyancy profile at equilibrium minus the potential energy of the background buoyancy profile. The background potential energy remains constant for the adiabatic dynamics, but the irreversible convergence of the system towards the equilibrium state implies an irreversible increase of potential energy for the system. Several results on the cumulative mixing efficiency are obtained within this framework:
\begin{enumerate}
\item The cumulative mixing efficiency increases in proportion to the Richardson number in the limit of small Richardson number, whatever the background buoyancy profile.
\item The cumulative mixing efficiency tends to $0.25$ in the limit of infinite Richardson numbers, provided that the background buoyancy profile is strictly decreasing with height (no homogeneous layer). 
\item The variations of the cumulative mixing efficiency with the Richardson number depends strongly on the background buoyancy profile, and can be non-monotonic. In the particular case of a fluid with two homogeneous layers of different buoyancy, the theory predicts a bell-shape for the cumulative mixing efficiency as a function of the global Richardson number.
\end{enumerate}

{  The application of equilibrium statistical mechanics to mixing in stratified fluids however relies on two key hypothesis:
\begin{enumerate}
\item  The theory applies to inviscid, adiabatic Boussinesq fluids. It is expected to describe mixing only in the limit of high Reynolds and Peclet numbers.
\item Equilibrium statistical mechanics relies on the counting of the available microscopic states, and its predictive power depends on the capability of the system to actually explore those available states.  Such an ergodic behaviour is favoured by stirring at the system scale $H$. By contrast turbulence forced at small scale $L_t$ with $L_t/H < 1$ is expected to produce  local mixing before large scale stirring, leading to discrepancies of the statistical mechanics predictions.
\end{enumerate}}

Note finally, the theory does not predict how the system converges towards equilibrium.  It does neither predict a turbulent diffusivity nor a turbulent viscosity during the relaxation process, but only the final outcome of turbulent stirring. And it does not account for energy fluxes in forced-dissipative configurations.

Equilibrium statistical mechanics therefore describes an ideal state of inviscid stirring which is not fully reached in most cases. Turbulent stirring could be however modelled locally as a trend to approach this equilibrium. This can be done by giving  a dynamical meaning to the relaxation equations used in this paper as an algorithm to compute the equilibrium state. Indeed, those equations contain a classical term modelling turbulent transport as an effective diffusion, with an additional drift term describing restratification. We believe that this approach will be fruitful to model relaminarisation after a mixing event, see e.g. \cite{venaillesommeria2011}. This would provide a complementary point of view to other statistical or stochastic approaches that have long been used in the context of combustion  \citep{pope1985pdf}, and adapted to the case of turbulent mixing in stratified fluids  \cite{kerstein1999one}. We hope the present paper will motivate further studies in those directions.\\

We warmly thank T. Dauxois, C. Herbert, P. Odier and  C. Staquet for useful discussions. Part of this work has been funded by ANR-13-JS09-0004-01 (STRATIMIX).}

\appendix
{ 
\section{Liouville theorem \label{app:Liouville}}

We show in this appendix that the quadruplet of fields  $(\mathbf{u},b)$ satisfy a Liouville theorem, i.e. that trajectories of the system are non-divergent in a phase-space described by this quadruplet of fields. The fact that Fourier components of the velocity field in each direction satisfy a detailed Liouville theorem is a classical result for three-dimensional Euler dynamics \citep{lee52}. Generalisation of this results to the inviscid, adiabatic Boussinesq system is straightforward, but is reproduced here for completeness. Let us for that purpose decompose both  the velocity field and the buoyancy field on Fourier modes:
\begin{equation}
\mathbf{u}=\sum_{\mathbf{k}}\hat{\mathbf{u}}_{\mathbf{k}}(t)e^{i\mathbf{k}\cdot\mathbf{x}},\quad b=\sum_{\mathbf{k}}\hat{b}_{\mathbf{k}}(t)e^{i\mathbf{k}\cdot\mathbf{x}}.\label{eq:fourier_def-1}
\end{equation}
Writing $\mathbf{u}=(u_{1},u_{2},u_{3})$ and projecting the equations of motion (\ref{eq:incompressibility})-(\ref{eq:advection_buoyancy})-(\ref{eq:momentum}) on a mode $\mathbf{k}$ yields 
\begin{eqnarray}
\dot{\hat{b}}_{\mathbf{k}} &=&-\mbox{i}\sum_{\mathbf{p}+\mathbf{q}=\mathbf{k}}\left(\hat{\mathbf{u}}_{\mathbf{p}}\cdot\mathbf{q}\right)\hat{b}_{\mathbf{q}}\label{eq:buoyancy_adv_fourier-1}, \\
\dot{\hat{u}}_{i\mathbf{k}}&=&\sum_{j,l}\left[\left(\delta_{i3}-\frac{k_{3}k_{i}}{k^{2}}\right)\hat{b}_{\mathbf{k}}+\left(\frac{k_{i}k_{j}}{k^{2}}-\delta_{ij}\right)\sum_{\mathbf{p}+\mathbf{q}=\mathbf{k}}q_{l}\hat{u}_{l\mathbf{p}}\hat{u}_{j\mathbf{q}}\right].\label{eq:momentum_fourier_modes_projected_cartesian_coord-1}
\end{eqnarray}
The pressure term has been eliminated from the momentum equation by using the non-divergence condition. Deriving Eq. (\ref{eq:buoyancy_adv_fourier-1}) by $b_{\mathbf{k}}$ and  Eq. (\ref{eq:momentum_fourier_modes_projected_cartesian_coord-1}) by $u_{i\mathbf{k}}$ allows us to show the existence of a detailed Liouville theorem for the Fourier components of the buoyancy field $b$, and for the Fourier components of the velocity field in each direction: 
\begin{equation}
\forall\mathbf{k},\quad\frac{\partial\dot{b}_{\mathbf{k}}}{\partial \hat{b}_{\mathbf{k}}}+\frac{\partial\dot{\hat{b}}_{-\mathbf{k}}}{\partial \hat{b}_{-\mathbf{k}}}=0,\quad\mbox{and }\forall i,\mathbf{k}\quad\frac{\partial\dot{\hat{u}}_{i\mathbf{k}}}{\partial \hat{u}_{i\mathbf{k}}}+\frac{\partial\dot{\hat{u}}_{i-\mathbf{k}}}{\partial \hat{u}_{i-\mathbf{k}}}=0 .\label{eq:detail_liouville_b-1}
\end{equation}
Using $(u_{1},u_{2},u_{3})=(u,v,w)$, we conclude that the quadruplet of fields $(u,v,w,b)$ satisfies a Liouville theorem: 
\begin{equation}\sum_{\mathbf{k}}\left[\frac{\partial\dot{\hat{b}}_{\mathbf{k}}}{\partial \hat{b}_{\mathbf{k}}}+\frac{\partial\dot{\hat{u}}_{\mathbf{k}}}{\partial \hat{u}_{\mathbf{k}}}+\frac{\partial\dot{\hat{v}}_{\mathbf{k}}}{\partial \hat{v}_{\mathbf{k}}}+\frac{\partial\dot{\hat{w}}_{\mathbf{k}}}{\partial \hat{w}_{\mathbf{k}}} \right]=0. \label{ref:liouville_theorem_app} \end{equation}

This Liouville theorem expresses the conservation of volume in the space of spectral amplitudes. However the discrete approximation of the fields that we propose in this paper relies on a uniform microscopic grid in physical space, and one needs to show that the Liouville property is not broken by this discrete approximation. We note for that purpose (i) that the Liouville property in Eq. (\ref{ref:liouville_theorem_app}) remains valid if the sum is truncated at wavenumbers $k_i\le N/2$ for $1\le i\le 3$, whatever the value of $N$, and (ii) that for a given truncation of the fields in Fourier space, the spectral amplitudes are related to the values of the fields on a collocation grid uniform in physical space, through a linear transformation that does no depend on the fields. The Jacobian of the transformation is therefore an unimportant constant, as noted in \cite{miller1990}. We conclude that a Liouville theorem holds for the finite-dimensional approximation of the buoyancy and velocity fields on a uniform grid.
}

\section{From Boltzmann entropy to macrostate entropy \label{app:A}}

{  The aim of this appendix is to count the number of microscopic configurations  $\mathbf{u}(\mathbf{x}),b(\mathbf{x})$ associated with a given macroscopic state $\rho(\mathbf{x},\sigma,\mathbf{v})$. In order to simplify the presentation, we show first how to count the number of microscopic configurations  $b(\mathbf{x})$ associated with a given macroscopic state $\rho(\mathbf{x},\sigma)$. The first step is to introduce a discrete approximation of the fields. The second step is a classical counting arguments within each macrocell of the discrete model. The third step is to consider the limit of an infinite number of grid point within each macrocell, which corresponds to the continuous limit for the microscopic configurations. The last step is to consider the limit of an infinite number of macrocells, which corresponds to the continuous limit for the macroscopic states, or equivalently  to the limit of a vanishing coarse-graining length scale.}

We assume that the domain $\mathcal{V}_{\mathbf{x}}$ is divided into a uniform grid containing $N$ cubic macrocells indexed by $1 \le I \le N$, and that each macrocell is divided into another uniform grid containing $M$ sites, where each site contains one and only one fluid particle indexed by $1 \le i \le M$, see Fig. \ref{fig:schema}. 
We also assume that the buoyancy ${b_{I,i}}$  at site $(I,i)$ can only take on a discrete number of values (say $K$), with $b_{I,i} \in \{\sigma_1,...,\sigma_K \}$, and that each of the resulting microstates is equiprobable. 
{  We note that with this procedure, we count the fields that will not be differentiable when taking the continuous limit, and we will see that the equilibrium state is actually characterised by such states containing  fluctuations of buoyancy at scale smaller than a coarse-grained scale, no matter how small the coarse-grained scale is.}
For a given discretised buoyancy field, we call $M_{I,k}$ the number of fluid particles carrying the buoyancy level $\sigma_k$ within the macrocell $I$, and $M_{I,k}/M$ is therefore the frequency of occurrence of the level $\sigma_k$ at site $I$ for one realisation of the discretised field.   
The system is described at a macroscopic level by the probability $p_{I,k}$ of measuring the buoyancy level $\sigma_k$ at site $I$. 

Our aim is to count number of microscopic configurations associated with a prescribed field $p_{I,k}$. 
We use for that purpose the equivalence between probability and frequency in the large $M$ limit:
\begin{equation}
p_{I,k}=\lim_{M\rightarrow +\infty} \frac{M_{I,k}}{M} \ .
\end{equation}
In the large $M$ limit, the number of microscopic discretised buoyancy fields $\{b_{I,i}\}_{1\le I \le N,1 \le i \le M }$ associated with the macroscopic field $\{p_{I,k}\}_{1\le I \le N,1 \le k\le K }$ is 
\begin{equation}
\Omega =\prod_{I=1}^N \left(\frac{M!}{\prod_{k=1}^K \left(M p_{I,k} \right)!} \right)\ . 
\end{equation}

The Boltzmann entropy is defined as 
\begin{equation}
S_{B}= k_B \log \Omega\ ,
\end{equation}
where $k_B$ is a constant.  
In the large $M$ limit, Stirling formula ($\log M!= M\log M$) leads at lowest order to 
\begin{equation}
S_{B}= -k_B M \sum_{I=1}^N \sum_{k=1}^K p_{I,k} \log p_{I,k} ,\label{eq:S_discret}
\end{equation}
where we have kept only the dominant term, and removed an unimportant constant depending on the grid size $M$.\\

It is important to note that for a given macrostate $\{p_{I,k}\}_{1\le I \le N,1\le k \le K }$, the number of possible microscopic  configurations $\Omega$ diverges exponentially with $M$, which a coefficient given by $ -\sum_{I=1}^N \sum_{k=1}^K p_{I,k} \log p_{I,k}  $. 
This means that among a set of different macrostates, there will be an overwhelming number of microstates associated with the one that maximises the coefficient  $ -\sum_{I=1}^N \sum_{k=1}^K p_{I,k} \log p_{I,k}  $. 
In other words, a single microscopic configuration picked up at random has a very large probability of being close to the macroscopic equilibrium state.
A practical consequence of this \emph{concentration property} is that no particular average procedure is required to observe the actual macroscopic equilibrium state. 
\\

In the limit $N \rightarrow +\infty $, the sum over $I$ in Eq. (\ref{eq:S_discret}), can be replaced by an integral over the spatial coordinate $\mathbf{x}$ if the discretised probability field $\{p_{I,k}\}_{1\le I \le N,1\le k \le K}$ is also replaced by its continuous counterpart $\{p_K \left(\mathbf{x} \right)\}_{1\le k \le K}$:
\begin{equation}
S_{B}= -k_B \frac{M N}{V}   \int_{\mathcal{V}_{\mathbf{x}}} \mathrm{d}\mathbf{x} \  \sum_{k=1}^K p_{k} (\mathbf{x}) \log p_{k} (\mathbf{x}) ,\label{eq:S_disctreteK,continuumN}
\end{equation}
Note that the quantity $p_{k} (\mathbf{x})$ is normalised at each point $\mathbf{x}$, with $\sum_{k=1}^K p_{k} (\mathbf{x})=1 $. It describes the local fluctuations of the (continuous) microscopic field $b$ in the vicinity of point $\mathbf{x}$, and it is called a \emph{Young measure} in mathematics.
A generalisation to the case of a continuum of  buoyancy levels $\sigma\in \mathcal{V}_{\sigma}=\left[\sigma_{min}\ \sigma_{max}\right]$ with probability density function  $\rho(\sigma,\mathbf{x})$ is less straightforward and requires the use of Sanov's theorem, see e.g. \cite{touchette2009large}. 
However, the result is easily inferred  from Eq. (\ref{eq:S_disctreteK,continuumN}) by decomposing the interval $\left[\sigma_{min}\ \sigma_{max}\right]$ into $K$ levels $\sigma_k$ equally spaced with interval $\Delta \sigma$, and by considering $\rho(\mathbf{x},\sigma_k) = p_k(\mathbf{x})/\Delta \sigma$.  Taking the limit $K\rightarrow +\infty $ yields 
\begin{equation}
S_{B}= -k_B \frac{M N}{V}   \int_{\mathcal{V}_{\mathbf{x}}} \mathrm{d}\mathbf{x}   \int_{\mathcal{V}_{\sigma}} \mathrm{d}\sigma \    \rho\left(\mathbf{x},\sigma\right) \log \rho\left(\mathbf{x},\sigma\right) ,\label{eq:S_continuum_b}
\end{equation}
up to an unimportant term depending on $K$. 
The quantity  $\rho (\mathbf{x},\sigma)$ is now the probability density function of measuring the buoyancy level $b=\sigma$ at height $z$, with the normalisation constraint $\int_{\mathcal{V}_{\sigma}} \mathrm{d} \sigma \ \rho(\mathbf{x},\sigma)=1$.\\

We are now ready to generalise this result to the case where a fluid particle at point $\mathbf{x}$ is carrying not only a buoyancy level $b(\mathbf{x})=\sigma$ with $\sigma \in \mathcal{V}_{\sigma}$, but also a velocity (vector) level $\mathbf{u}(\mathbf{x})=\mathbf{v}$ with $\mathbf{v} \in \mathcal{V}_{\mathbf{v}}=\left[-v_{max},\  v_{max}\right]^3 $.  
The same steps leading to Eq. (\ref{eq:S_continuum_b}) can be applied to that case, which yields
\begin{equation}
S_{B}= -\frac{k_B M N}{V} \int_{\mathcal{V}_{\mathbf{x}}} \mathrm{d} \mathbf{x} \int_{\mathcal{V}_{\sigma}} \mathrm{d} \sigma \int_{\mathcal{V}_{\mathbf{v}}} \mathrm{d} \mathbf{v} \ \rho \left(\mathbf{x},\sigma,\mathbf{v}\right) \log \rho\left(\mathbf{x},\sigma,\mathbf{v}\right) \label{eq:S_continuum_levels}
\end{equation}
where $\rho \left(\mathbf{x},\sigma,\mathbf{v}\right)$ is the probability density function for the buoyancy and velocity at point  $\mathbf{x}$.
Note that we have introduced a cut-off denoted $v_{max}$  for the maximum possible velocity. 
Anticipating that velocity fluctuations are bounded due to the energy constraint, we expect that the results will not depend on $v_{max}$ if it is chosen much larger than the root mean square velocity of the equilibrium state, and we will consider in the remaining of this paper $v_{max}=+\infty$.

Finally, choosing $k_B=V/(N M)$ in Eq. (\ref{eq:S_continuum_levels}), we recover $S_B=\mathcal{S}[\rho]$, where  $\mathcal{S}[\rho]$ is the macrostate entropy defined in Eq. (\ref{eq:mixing_entropy}).

\section{Relaxation equations from a maximum entropy production principle \label{app:B}}
{  The aim of this appendix is to provide an algorithm that makes possible numerical computations of the equilibrium states for arbitrary energy $E$ and global distribution of buoyancy  $G(\sigma)$. We consider for that purpose the ansatz (\ref{eq:ansatz}) for the local distribution of  velocity and buoyancy levels, and we propose in the following a dynamical system describing the temporal evolution of the quantities  $\rho_b(\mathbf{x},\sigma,t)$, $e_c(t)$ in such a way that the total energy and the global distribution of buoyancy levels are conserved, just as in the original Boussinesq system,  and in such a way that the entropy production is maximum at each time. This maximum entropy production principle ensures convergence towards an entropy maximum for a given set of constraints $E$, $G(\sigma)$.  Since the effective temperature (i.e. the Lagrange parameter associated with the energy) is positive, the entropy maximum is unique for a given set of constraints, and the dynamical system will therefore relax towards the equilibrium state. We stress that considering the temporal evolution of this dynamical system is a trick to find the equilibrium state. The actual flow dynamics may follow a different path towards equilibrium than the one maximizing the entropy production.} 

Since the dynamical system is fully described by $\rho_b(\mathbf{x},\sigma,t)$ and $e_c(t)$, it will be useful in the following to express the conservation of the global buoyancy distribution and of the total energy in terms of those parameters. Inserting  Eq. (\ref{eq:ansatz}) in (\ref{eq:HHH}) and (\ref{eq:E_rhop}) yields to 
\begin{equation}
\mathcal{G}[\rho_b] = \frac{V}{2H} \int_{-H}^{H} \mathrm{d} z  \ \rho_b  \ .\label{eq:HHHApp}
\end{equation}
\begin{equation}
\mathcal{E}[\rho_b](e_c)= V e_c +\frac{V}{2H}\int_{-H}^{H} \mathrm{d} z  \int_{\mathcal{V}_{\sigma}}  \mathrm{d}\sigma \ \rho_b \sigma z \ .\label{eq:E_rhopApp}
\end{equation}
If the initial condition $\rho_b(z,\sigma,0)$ and the initial kinetic energy $e_c(0)$ are known, then the global distribution of buoyancy levels and total energy can be computed using Eq. (\ref{eq:HHHApp}) and Eq. (\ref{eq:E_rhopApp}), respectively.

Assuming that there is no source nor sink of density, {  recalling that the flow is non-divergent},  and anticipating that there is no mean flow, the temporal evolution of the pdf $\rho_b$ satisfies the general conservation law
\begin{equation}
\partial_t \rho_b + \partial_z  J_b =0\ , \label{eq:general_conservation_law}
\end{equation}
where we have introduced the turbulent flux of probability $J_b(z,\sigma,t )$ directed along $z$, with $J_b=0$ at the upper and the lower boundary $z=\pm H$. 

The temporal evolution of the system requires a model for the flux  $J_b$ and  the kinetic energy production $\dot e_c =\mathrm{d} e_c/\mathrm{d} t$.
The maximum entropy production principle amounts to finding the flux $J_b$ and the kinetic energy production $\dot e_c$ that maximise the entropy production while satisfying the constraints of the problem.

Let us first compute the entropy and energy production. Injecting the ansatz (\ref{eq:ansatz}) in Eq. (\ref{eq:mixing_entropy}), the macrostate entropy  can be expressed as 
\begin{equation}
\mathcal{S}= - \frac{V}{2H} \int_{-H}^{+H} \mathrm{d}z \int_{\mathcal{V}_{\sigma}}  \mathrm{d}\sigma \ \rho_b \log \rho_b   - \frac{3}{2}  V  \log \frac{3}{2e_c}  \label{eq:entro0} .
\end{equation}
Taking the time derivative of Eq. (\ref{eq:entro0}) and using Eq. (\ref{eq:general_conservation_law}), the entropy production can be expressed as  
\begin{equation}
\dot \mathcal{S}=  \frac{V}{2H}  \int_{-H}^{+H} \mathrm{d}z \int_{\mathcal{V}_{\sigma}}  \mathrm{d}\sigma \ J_b \partial_z (\log \rho_b) \  - \frac{3}{2} V \frac{\dot e_c}{e_c}. \label{eq:entro}
\end{equation}
The constraints of the problem are given by
\begin{itemize}
\item the conservation of the local normalisation  (\ref{eq:norm}), implying  
\begin{equation}
\forall z \in [-H\  H ],\  \int_{\mathcal{V}_{\sigma}} \mathrm{d} \sigma \ J_b =0, \label{eq:constraintJ1}
\end{equation}
which ensures the local normalization $\int_{\mathcal{V}_{\sigma}} \mathrm{d} \sigma \rho_b =1$,
\item {  the energy conservation, which can be expressed as $\dot{\mathcal{E}}=0$. Taking the temporal derivative of Eq. (\ref{eq:E_rhopApp}) yields to }
\begin{equation}
\dot \mathcal{E}=V \dot e_c+ \frac{V}{2H}\int_{-H}^{+H} \mathrm{d} z  \int_{\mathcal{V}_{\sigma}}\mathrm{d} \sigma  \ \sigma  J_b   .\label{eq:constraintEnergyProd}
\end{equation}
\item {  Finally, the fluxes of probability must be finite to be dynamically relevant. Indeed, an infinite flux would corresponds to an instantaneous rearrangements of the buoyancy field. We impose therefore a bound for the norm of the probability flux $J_b$, expressed as:}
\begin{equation}
\forall z \in [-H\  H ],\   \ \int_{\mathcal{V}_{\sigma}} \mathrm{d} \sigma\  \frac{J_b^2}{2\rho_b} \le C(z) \ . \label{eq:constraintJ2}
\end{equation}
The quantity $J_b / \rho_b$ can be interpreted as a diffusion velocity for the probability density field, and the constraint in Eq. (\ref{eq:constraintJ2}) ensures that this velocity remains finite everywhere and for each buoyancy level during the relaxation process.
\end{itemize}

The variational problem of the maximum entropy production principle is treated by introducing Lagrange multipliers $\zeta(z)$, $\beta_t$ and $-/ D(z)$ associated with  the constraints in Eqs. (\ref{eq:constraintJ1}), (\ref{eq:constraintEnergyProd}) and  (\ref{eq:constraintJ2}), respectively. {  Note that following the Karush-Kuhn-Tucker conditions, an inequality such as (\ref{eq:constraintJ2})  can be treated as an equality constraint when computing the first order variations in an optimisation problem \citep{sundaram1996first}.} 
The condition
\begin{equation}
\delta\dot \mathcal{S} - \beta_t \delta \dot \mathcal{E} +\int_{-H}^{+H} \mathrm{d} z\ \int_{\mathcal{V}_{\sigma}} \mathrm{d}\sigma \ \zeta(z)\delta J_b- \int_{-H}^{+H} \mathrm{d} z\ \int_{\mathcal{V}_{\sigma}} \mathrm{d}\sigma \ \frac{1}{D}\frac{J_b}{\rho_b}\delta J_b =0\ ,\label{eq:prod_entr_frist_var}
\end{equation}
must be satisfied for each $\delta \dot e_c$ and $\delta J_b$. Using  
\begin{equation}
\delta\dot \mathcal{S} =\frac{V}{2H}\int_{-H}^{+H} \mathrm{d}z \int_{\mathcal{V}_{\sigma}}  \mathrm{d}\sigma \   \partial_z (\log \rho_b) \delta J_b \  - \frac{3V}{2e_c}  \delta \dot e_c ,\ \ \delta \dot \mathcal{E} =V \delta \dot e_c+ \frac{V}{2H}\int_{-H}^{+H} \mathrm{d} z  \int_{\mathcal{V}_{\sigma}}\mathrm{d} \sigma  \ \sigma  \delta J_b , \nonumber
\end{equation}
Eq. (\ref{eq:prod_entr_frist_var}) yields 
\begin{equation}
\beta_t=3/(2e_c), \quad J_b=-D\left(\partial_{z}\rho_b-\beta_{\mathrm{t}} \left(\sigma-\overline{b} \right) \rho_b \right), \label{eq:Jrelaxation}
\end{equation}
where $\zeta(z)$ has been determined by using the constraint in Eq. (\ref{eq:constraintJ1}). In addition, the coefficient $D$ must be positive for the entropy production to be positive. As far as the equilibrium state is concerned, the value of $D$ in not important. Indeed, the flux $J_b$ vanishes when equilibrium is reached, which ensures that the equilibrium state does not depend on $D$.

\newpage

\bibliographystyle{jfm}

\bibliography{twolayers}

\end{document}